\documentclass[10pt,twocolumn,letterpaper]{IEEEtran}
\usepackage[T1]{fontenc}
\usepackage[latin1]{inputenc}
\usepackage{color}
\usepackage{psfrag}
\usepackage[dvips]{graphicx}
\usepackage{tikz}
\usepackage{pgfplots}
\pgfplotsset{compat=1.12}
\usepackage{enumerate}
\usepackage{rotating}
\usepackage{srcltx}
\usepackage{booktabs}
\usepackage{enumitem}
\usetikzlibrary{arrows,decorations.pathmorphing,backgrounds,positioning,fit,matrix}

\usepackage[noadjust]{cite}
\usepackage[cmex10]{amsmath}
\usepackage{amsfonts}
\usepackage{amssymb}
\usepackage{ntheorem}
\usepackage{dsfont}    
\usepackage{mathtools} 
\usepackage{stackengine}

\definecolor{PU_orange}{HTML}{EE7F2D}
\definecolor{PU_darkorange}{HTML}{994400}
\definecolor{PU_lightorange}{HTML}{FFAA66}
\definecolor{PU_black}{HTML}{000000}
\definecolor{PU_darkgray}{HTML}{7F7F83}
\definecolor{PU_lightgray}{HTML}{BDBEC1}

\usepackage{algorithmic}
\usepackage{array}
\usepackage[caption=false,font=footnotesize]{subfig}

\theoremseparator{.}
\newtheorem{corollary}{Corollary}
\newtheorem{prop}{Proposition}
\newtheorem{lem}{Lemma}
\newtheorem{theorem}{Theorem}
\theoremseparator{.}
\newtheorem{definition}{Definition}
\newtheorem{remark}{Remark}
\newtheorem*{example*}{Example}

\newcommand{\eu}{\mathrm{e}}

\newcommand{\mmse}{\mathrm{mmse}}

\newcommand{\E}{\mathbb{E}}

\long\def\symbolfootnote[#1]#2{\begingroup%
	\def\thefootnote{\fnsymbol{footnote}}\footnote[#1]{#2}\endgroup} 
\title{A Cram\'er-Rao Type Bound for Bayesian Risk with Bregman Loss}%

\author{ 
    Alex~Dytso,~\IEEEmembership{Member,~IEEE,} 
    Michael~Fau\ss,~\IEEEmembership{Member,~IEEE,}
    and~H.~Vincent~Poor,~\IEEEmembership{Life~Fellow,~IEEE}
    \thanks{This work was supported by the U.S. National Science Foundation under Grant CCF-1908308.}
    \thanks{The work of M.~Fau\ss{} was supported by the German Research Foundation (DFG) under grant number 424522268.}%
    \thanks{A.~Dytso, M.~Fau\ss, and H.~V.~Poor are with the Department of Electrical Engineering, Princeton University, Princeton, NY 08544, USA. E-mail:  adytso@prinction.edu, mfauss@princeton.edu, poor@princeton.edu.}
}
\IEEEoverridecommandlockouts
\begin{document}
\maketitle

\begin{abstract}
A general class of Bayesian lower bounds  when the underlying loss function is a  Bregman divergence is demonstrated. This class can be considered as an extension of the Weinstein--Weiss family of bounds for the mean squared error and relies on finding a variational characterization of Bayesian risk.    The approach allows for the derivation of a version of the Cram\'er--Rao bound that is specific to a given Bregman divergence. The new generalization of the Cram\'er--Rao bound reduces to the classical one when the loss function is taken to be the Euclidean norm. 
The effectiveness of the new bound is evaluated in the Poisson noise setting and the Binomial noise setting.  
\end{abstract}

\section{Introduction} 


Finding lower bounds on \emph{a Bayesian risk} is an important aspect in signal estimation as such bounds provide fundamental limits on signal recovery. Moreover, they can contribute useful insights and guidelines for algorithm design in data-driven applications, where Bayesian analysis is of ever-increasing importance.  A plethora of such bounds are known for the mean squared error  (MSE). Loosely speaking, these bounds can be divided into three families. The first family, termed \emph{Weinstein--Weiss}, works by using the Cauchy--Schwarz inequality \cite{weinstein1988general}, and includes the prevalent \emph{Cram\'er--Rao} (CR) bound (also known as the van-Trees bound \cite{van2004detection})  as a special case.  The second family,  termed \emph{Ziv--Zakai}, is derived by connecting estimation and binary hypothesis testing \cite{ziv1969some}. The third family uses a  variational approach and works by minimizing the MSE subject to a constraint on a suitably chosen divergence measure, for example, the \emph{Kullback--Leibler} divergence \cite{MMSEKLpreparation,MMSEJointArxiv}. 

This paper is concerned with a generalization of the Weinstein--Weiss family of bounds beyond the MSE.  Specifically, the aim is to provide a generalization of this family to a larger class of Bayesian risks, where the loss functions are taken to be \emph{Bregman divergences} (BDs).  The difficulty with such a generalization is that  it is not immediately clear how the Cauchy--Schwarz inequality can be applied to a Bregman divergence, which is, in general, not a metric and, at a first glance, does not have a natural norm associate with it. 

In this paper, by using elementary techniques such as Taylor's remainder theorem, it is shown that the Weinstein--Weiss approach can be generalized to the Bayesian risks where the loss function is taken to be a BD. Furthermore, this generalization makes it possible to derive a version of the CR bound that is specific to a given BD. The new generalization of the CR bound reduces to the classical CR bound when the loss function is taken to be the Euclidean norm.  

The paper outline and contributions are as follows: 
\begin{itemize}[leftmargin=*]
    \item Section~\ref{sec:Bregman Divergence:Basics} reviews properties of Bregman divergences and of the corresponding Bayesian risk. 
    \item Section~\ref{sec:VariaationalRepresentationOfBregmanRisk} presents a variational characterization of the Bregman risk and the new family of bounds. 
    \item Section~\ref{sec:ApplicationsOfVar} discusses  some applications of the variational representation.   In particular,  Theorem~\ref{thm:CRboundBayesian} presents a generalization for the CR bound that is specific to a given BD and reduces to the classical CR bound when the Bayesian risk corresponds to the MSE.
    \item Section~\ref{sec:CRboundsEvaluation}, in order to show the utility of the new CR bound, evaluates the CR bound for the Poisson noise case with a BD natural for this setting.   In particular, it is shown that the CR bound has the same behavior as the Bayesian risk when the scaling parameter of the Poisson noise is taken to be large. 
    \item Section~\ref{sec:BinamialSection} evaluates the new CR bound for a second example, namely for a Binomial noise with a BD natural for this setting. In the Binomial noise case, similarly to the Poisson noise setting, the CR bound is shown to be order tight in a certain regime. 
    \end{itemize}

\paragraph*{Notation} Random variables are denoted by upper case letters, and their realizations are denoted by lower case letters.  The inner product operator is denoted by $\langle \cdot, \cdot \rangle$.  The identity matrix is denoted by $\mathsf{I}$.  For two symmetric matrices $\mathsf{A}$ and $\mathsf{B}$ we say that $\mathsf{A}\prec \mathsf{B}$ if  $\mathsf{B}-\mathsf{A}$ is positive-definite.  For    $\mathsf{0}\prec \mathsf{A}$ we define the Mahalanobis metric as $\|x\|_{\mathsf{A}}=x^T \mathsf{A} x$, where $\| x \|$ denotes the Euclidian mectric.   The expected value operator is denote by $\E[ \cdot]$. For a random variable $X \in \mathbb{R}^n$ with a probability density function (pdf) $f_X$, the score function is defined as $\rho_X(x)=\frac{\nabla f_X(x)}{f_X(x)}$, where $\nabla$ is the gradient operator.

\section{Bregman Divergence and Bayesian Risk}
\label{sec:Bregman Divergence:Basics}

In order to define a Bayesian risk or estimation error, one needs to select a loss function. The family of loss functions considered in this paper is defined next. 

\begin{definition}    \label{def:bregman_divergence} \emph{(Bregman Divergence.)}  
    Let $\phi: \Omega \to \mathbb{R}$ be a \emph{continuously-differentiable} and \emph{strictly convex}  function defined on a \emph{closed convex} set $\Omega \subseteq \mathbb{R}^n$.   The Bregman divergence between $u$ and $v$ associated with the function $\phi$ is defined as
    \begin{align}
        \ell_\phi(u,v) =\phi(u)-\phi(v)-\langle u-v, \nabla \phi(v) \rangle.  \label{eq:Definition:Bregman:Divergence}
    \end{align} 
\end{definition}

\begin{figure}[tb]  
	\centering   
%
%
\definecolor{mycolor1}{rgb}{0.00000,0.44700,0.74100}%
\definecolor{mycolor2}{rgb}{0.85000,0.32500,0.09800}%
\begin{tikzpicture}
\small
\begin{axis}[%
width=\columnwidth,
height=0.7\columnwidth,
xmin=0.502880184331797,
xmax=2.63882488479263,
ymin=-0.204379562043796,
ymax=5,
xlabel={$u$},
axis background/.style={fill=white},
xmajorgrids,
ymajorgrids,
legend style={legend cell align=left, align=left, draw=white!15!black}
]
\addplot [color=black,thick]
  table[row sep=crcr]{%
0	0\\
0.1	0.01\\
0.2	0.04\\
0.3	0.09\\
0.4	0.16\\
0.5	0.25\\
0.6	0.36\\
0.7	0.49\\
0.8	0.64\\
0.9	0.81\\
1	1\\
1.1	1.21\\
1.2	1.44\\
1.3	1.69\\
1.4	1.96\\
1.5	2.25\\
1.6	2.56\\
1.7	2.89\\
1.8	3.24\\
1.9	3.61\\
2	4\\
2.1	4.41\\
2.2	4.84\\
2.3	5.29\\
2.4	5.76\\
2.5	6.25\\
2.6	6.76\\
2.7	7.29\\
2.8	7.84\\
2.9	8.41\\
3	9\\
};

\addplot [color=black,dashed, thick]
  table[row sep=crcr]{%
0	-1\\
0.1	-0.8\\
0.2	-0.6\\
0.3	-0.4\\
0.4	-0.2\\
0.5	0\\
0.6	0.2\\
0.7	0.4\\
0.8	0.6\\
0.9	0.8\\
1	1\\
1.1	1.2\\
1.2	1.4\\
1.3	1.6\\
1.4	1.8\\
1.5	2\\
1.6	2.2\\
1.7	2.4\\
1.8	2.6\\
1.9	2.8\\
2	3\\
2.1	3.2\\
2.2	3.4\\
2.3	3.6\\
2.4	3.8\\
2.5	4\\
2.6	4.2\\
2.7	4.4\\
2.8	4.6\\
2.9	4.8\\
3	5\\
};

\node[below left, align=left]
at (axis cs:2,4.6) {$\phi(u)$};

\node[below right, align=left]
at (axis cs:0.81,1.6) {$\phi(v)$};
\node[below right, align=left]
at (axis cs:2,4.1) {$\ell_{\phi}{(u,v)}$};

\draw[>=triangle 45, <->] (2,4) -- (2,3);
\end{axis}

\end{tikzpicture}%
	\caption{Illustration of the definition of the Bregman divergence.}
	\label{fig:BregmanIllustration}
\end{figure}
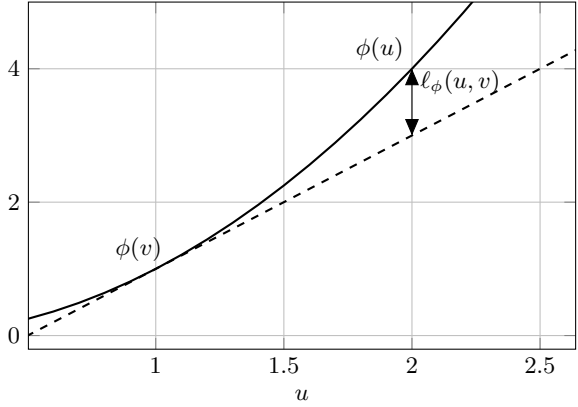

 The BD can be interpreted as an error due an approximation of $\phi(u)$ with a line tangent to the point $(v,\phi(v))$. 
Fig.~\ref{fig:BregmanIllustration} illustrates this interpretation. 
  
BDs have been introduced in \cite{bregman1967relaxation} in the context of convex optimization. There  exists several extension of the BD definition such as  an extension to functional spaces \cite{frigyik2008functional},  an extension to submodular set functions  \cite{iyer2012submodular}, and a matrix extension \cite{wang2014bregman}.   

In \cite{csiszar1991least}, BDs, together with $f$-divergences, where characterized axiomatically. A thorough investigation of BDs was undertaken in \cite{banerjee2005clustering}, where it was shown that many commonly used loss functions are members of this family. Moreover,  the authors of \cite{banerjee2005clustering} have shown that every regular exponential distribution has a unique BD associated with it.  

Now consider the problem of estimating a random variable $X$ from a noisy observation $Y$, where the loss function is according to \eqref{eq:Definition:Bregman:Divergence}.  The smallest Bayesian risk associated with this estimation problem is defined next. 

\begin{definition} \emph{(Minimum Bayesian Risk with Respect to BD.)} 
    For a joint distribution $P_{Y,X}$ we denote the minimum Bayesian risk with respect to the loss function $\ell_\phi(u,v) $ as 
    \begin{align}
        R_\phi( X|Y ) =\inf_{f:  f \text{ is measurable}} \E \left[  \ell_\phi \left(X,f(Y) \right)  \right].  \label{eq:BregmanRisk}
    \end{align}
    $R_\phi( X|Y )$ is also referred to as \emph{Bayesian Bregman risk} in what follows.
\end{definition} 

\begin{remark}
    The most prominent example of $R_\phi( X|Y ) $ is the minimum mean squared error (MMSE), which is induced by choosing $\phi(u)= \|u\|^2$ and  will be denoted by 
    \begin{align}
        \mmse(X|Y)=R_\phi( X|Y ). 
    \end{align}
\end{remark}

The structure of the optimal estimator in \eqref{eq:BregmanRisk} was studied in \cite{banerjee2005optimality}, where it was shown that the conditional expectation is the unique minimizer. Moreover, the authors of \cite{banerjee2005optimality}  have also demonstrated the converse result, namely that the conditional expectation is an optimal estimator only when the loss function is a Bregman divergence.

\subsection{Fundamental Properties of Bregman Divergences} 
\label{sec:FThmBD-BR}

We now, for completeness, review the most important properties of BDs and the associated Bayesian risks. 

\begin{theorem}\label{thm:propBR-BD} 
  \emph{(Fundamental Properties of Bregman Divergences and Bayesian Bregman Risks.)}
    \begin{enumerate}[leftmargin=*]
        \item \emph{(Non-Negativity)}  $\ell_\phi(u,v) \ge 0, \forall  u,v \in \Omega$, with equality if and only if $u=v$;
        \item \emph{(Convexity)}  $\ell_\phi(u,v)$ is convex in $u$;
        \item  \emph{(Linearity)}  $\ell_\phi(u,v)$ is  linear in $\phi$;
        \item \emph{(Generalized Law of Cosines):}  For $u,v,w \in \Omega$ 
            \begin{align}
                &\ell_\phi(u,v) \notag\\
                &= \ell_\phi(u,w)+\ell_\phi(w,v)  - \langle u-w, \nabla \phi(v)- \nabla \phi(w)  \rangle;
            \end{align} 
        \item \emph{(Orthogonality Principle and Pythagorean Identity)}  For every random variable $X\in \Omega$ and every $u\in \Omega$ 
            \begin{align}
                \E \left[  \ell_\phi(X,u)\right]=   \E \left[  \ell_\phi(X,\E[X])\right]+ \ell_\phi(\E[X],u).
            \end{align}
            Moreover, for any measurable $f(Y)$
            \begin{align}
                &\E \left[  \ell_\phi(X,f(Y))\right] \notag\\
                &=   \E \left[  \ell_\phi \left(X,\E[X|Y] \right)\right]+  \E \left[ \ell_\phi \left(\E[X|Y],f(Y) \right) \right]. \label{thm:PythegorianIdentity}
            \end{align}
    \item \emph{(Conditional Expectation is the Unique Minimizer)}   Suppose that $\E[X]<\infty$ and $\E[\phi(X)]<\infty$. Then,
        \begin{align}
            \inf_{f:  f \text{ is measurable}} \E \left[  \ell_\phi \left(X,f(Y) \right)  \right]=   \E \left[  \ell_\phi \left(X, \E[X|Y] \right)  \right]. \label{eq:OptimalityOfCOnditionalExpectation}
        \end{align}
        The optimizer in \eqref{eq:OptimalityOfCOnditionalExpectation} is unique $Y$-almost surely;
        \item \emph{(Coupling between Conditional Expectation and Bregman Divergence)}   Let $F :\mathbb{R}^n \times  \mathbb{R}^n  \mapsto \mathbb{R}$ be a non-negative function such that $F(x,x)=0$ and assume that all partial derivatives $F_{x_i,x_j}$ are continuous.  If for all random variables $X \in \mathbb{R}^n$ it holds that
        \begin{align}
            \inf_{u \in \mathbb{R}^n}\E \left[  F(X,u)\right]=   \E \left[ F(X,\E[X])\right],
        \end{align}
        with $\E[X]$ being the unique minimizer, then $F(u,v)= \ell_\phi(u,v)$ for some strictly convex and differentiable function $\phi: \mathbb{R}^n \to \mathbb{R}$. 
    \end{enumerate} 
\end{theorem}

\subsection{Notable Examples of BDs in Estimation Theory}

BDs and their associated Bayesian risks appear naturally in connection with information measures. For example, the mutual information between $Y$ and $X$ can be represented as an integral of the BD induced by
\begin{itemize}[leftmargin=*]
    \item $\phi(u)=\|u\|^2$ (i.e., the MMSE) when $P_{Y|X}$ is a  Gaussian distribution \cite{I-MMSE,palomar2005gradient};
    \item $\phi(u)=u\log u$ when $P_{Y|X}$ is a  Poisson distribution \cite{guo2008mutual,atar2012mutual};
    \item $\phi(u)=u \log \frac{u}{1-u}$ when $P_{Y|X}$ is a  Binomial distribution \cite{taborda2014information}; and 
    \item $\phi(u)=u \log \frac{u}{1+u}$ when $P_{Y|X}$ is a  Negative Binomial distribution \cite{taborda2014information}.
\end{itemize} 
Table~\ref{table:BDexamples} summaries the above examples together with the corresponding BDs. The latter will be referred to as the \emph{natural} BDs in what follows. For a more detailed treatment of connections between information measures and BDs, the interested reader is referred to \cite{jiao2018mutual}. 

\begin{table}
    \centering
    \caption{Table of Bregman Divergences} \label{table:BDexamples}
    \begin{tabular}{@{}cccr@{}}
        \toprule
        Domain & $\phi( u ) $ & $\ell_\phi(u,v)$  &     Natural Noise  \\ 
        \addlinespace[1ex]
        \hline 
        \addlinespace[1.5ex]
        $ \mathbb{R}^n$   & $ \|u\|_{\mathsf{A}}^2$, \, $ \mathsf{0} \preceq \mathsf{A}$   & $\|u-v\|_{\mathsf{A}}^2$  &    Gaussian \\
        \addlinespace[1.5ex]
        $\mathbb{R}_+$ & $u \log u$  & $ u \log \frac{u}{v} - (u - v)$  &   Poisson \\
        \addlinespace[1.5ex]
        $[0,1]$ & $u \log \frac{u}{1-u}$  & $ u \log \frac{u(1-v)}{v(1-u)} - \frac{(u - v)}{1-v}$  &   Binomial \\
        \addlinespace[1.5ex]
        $\mathbb{R}_+$ & $u \log \frac{u}{1+u}$  & $ u \log \frac{u(1+v)}{v(1+u)} - \frac{(u - v)}{1+v}$  &  Negative Binomial \\
        \addlinespace[1ex]
        \bottomrule 
    \end{tabular}
    \vspace{-0.5cm}
\end{table}

\section{A Variational Representation of Bregman Risk}
\label{sec:VariaationalRepresentationOfBregmanRisk}
In this section, we provide a variational characterization of  Bayesian Bregman risk. We start by introducing an $\ell_2$-representation of Bregman divergences. 

\subsection{Bregman Divergence vs. the Mahalanobis distance} 
As one might expect, the variational characterization of the Bayesian Bregman risk requires an application of the Cauchy--Schwarz inequality. However,  a priori, it is not immediately clear how the Cauchy--Schwarz inequality can be applied to frequently complicated  expressions (see Table~\ref{table:BDexamples}) of  BDs.  The approach, however, becomes clear after an elementary application of Taylor's remainder theorem, which allows representing a BD as a weighted squared error. 
\begin{lem}\label{lem:BregmanRemeinder} \emph{($\ell_2$-representation of Bregman Divergences.)}
   Suppose that  $\phi$ in Definition~\ref{def:bregman_divergence}  is twice differentiable. Then, it holds that
    \begin{align}
      \ell_\phi(u,v) &=  \| (u-v)^T \mathsf{\Delta}_{\phi}^{\frac{1}{2}}(u,v)\|^2, \label{Eq:InterpretationOfBregman} 
    \end{align}
    where 
    \begin{align}
        \mathsf{\Delta}_\phi(u,v)=  \int_0^1 (1-t)   \mathsf{H}_{\phi}( (1-t)u+tv)  {\rm d} t  ,
        \label{eq:delta_phi}
    \end{align}
    and $x \mapsto  \mathsf{H}_{\phi} (x)$ is the Hessian matrix of $\phi$ evaluated at $x$. 
\end{lem}
\begin{IEEEproof}
Recall that given a twice differentiable  function  $\phi: \mathbb{R}^n \to \mathbb{R}$  Tailor's remainder theorem \cite{folland2005higher} asserts that
\begin{align}
    \phi(u)&=\phi(v)+ \langle u-v, \nabla \phi(v) \rangle  \notag\\
    &+ (u-v)^{T} \left[ \int_0^1 \! (1-t)   \mathsf{H}_{\phi}(u+t(v-u))  {\rm d} t \right]   (u-v), \label{eq:taylor_remainder}
\end{align} 
where  $x \mapsto  \mathsf{H}_{\phi} (x)$ is the Hessian matrix of $\phi$ evaluated at $x$. 

Observe that the BD $    \ell_\phi(u,v) $ is the remainder of the first order Tailor series expansion of $\phi(u)$ around $v$.  
  Therefore, by the integral representation of the Taylor expansion remainder in \eqref{eq:taylor_remainder}, it follows that
  \begin{align} 
    \ell_\phi(u,v) &= (u-v)^{T} \left[ \int_0^1 \! (1-t)   \mathsf{H}_{\phi}(u+t(v-u))  {\rm d} t \right]   (u-v) \\
    &= (u-v)^{T} \mathsf{\Delta}_\phi(u,v)   (u-v). \label{eq:bregman_weighted_square}
  \end{align}
  This concludes the proof. 
\end{IEEEproof}

\begin{remark} 
    In the scalar case \eqref{eq:bregman_weighted_square} simplifies to
    \begin{align}
        \ell_\phi(u,v) &=    \ell_{x^2}(u,v)  \Delta_\phi(u,v),
        \label{Eq:InterpretationOfBregmanScalar}
    \end{align}
    with $\Delta_\phi$ being strictly positive.
\end{remark}
\begin{remark} 
    It can be argued that \eqref{Eq:InterpretationOfBregman} and \eqref{Eq:InterpretationOfBregmanScalar} trivially hold true since any two functions with identical support can be expressed as weighted versions of one another by simply choosing the weight to be their ratio. Here, however, it is important to note that $\mathsf{\Delta}_\phi$ can be obtained directly from $\phi$ by evaluating the integral on the right hand side of \eqref{eq:delta_phi}. In other words, $\mathsf{\Delta}_\phi$ can be calculated without evaluating $\ell_\phi$.
\end{remark}

We note there were other attempts to connect $\ell_2$ distance and the BD. For instance, the  authors of \cite{li2012fast} also used the fact  that  BD   is the remainder of the first order Tailor series expansion and expressed the remained as an infinite series.  However, such infinite series representations require $\phi$ to be infinitely differentiable and do not lead to a compact representation as in \eqref{Eq:InterpretationOfBregman}.

\begin{figure}[tb]  
	\centering   
	\input{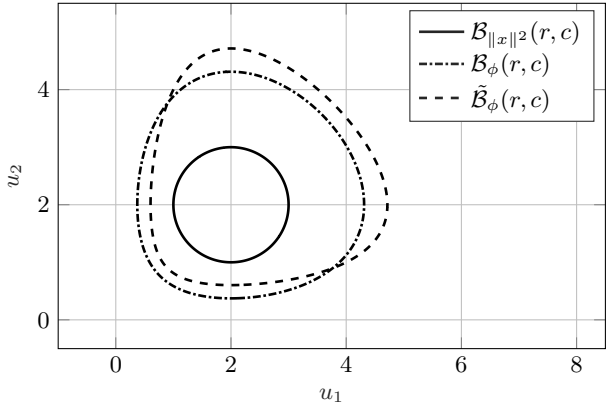}
	\caption{Comparison of Bregman balls $\tilde{\mathcal{B}}_{\phi}(r,c),\tilde{\mathcal{B}}_{\phi}(r,c)$ and the Euclidean ball $\mathcal{B}_{ \|  x \|^2}(r,c)$ where $r=1$ and $c=[2,2]$. }
	\label{fig:BregmanBall}
\end{figure}

Equation in \eqref{Eq:InterpretationOfBregman} has a strong resemblance to  Mahalanobis distance with the exception that the covariance matrix depends on the inputs $u, v$. That is,  using \eqref{Eq:InterpretationOfBregman}, we can write
\begin{align}
      \ell_\phi(u,v) = \|  u- v\|_{\mathsf{\Delta}_{\phi}(u,v)}^2.   \label{Eq:Mahalanobis_distance_Interpretation}
\end{align}
It is important to note that this analogy only holds \emph{locally}, meaning that at any given point $(u, v)$ the BD $\ell_\phi(u,v)$ corresponds to a certain Mahalanobis distance. However, since the latter changes with $(u, v)$, the \emph{global} properties of BDs and Mahalanobis distances, such as the expected risk considered here, can differ significantly.
For instance,  one can compare the topology of the ball induced by each divergence. To this end, let  the Bregman-ball of  radius $r$ with center at $c \in \Omega$ be defined as follows:
\begin{align}
\mathcal{B}_{\phi}(r,c)= \{  u \in \Omega:  \ell_{\phi}(u,c) \le r  \}.  
\end{align} 
Moreover, because $ \ell_{\phi}(u,v)$ may not by symmetric we can also define  $\tilde{\mathcal{B}}_{\phi}(r,c)= \{  u \in \Omega:  \ell_{\phi}(c,u) \le r  \}$. For the Mahalanobis distance, the shape of the ball or the neighborhood is given by an ellipse. This is not the case for BDs. For example, consider the function $\phi(u)=u_1 \log u_1 + u_2 \log u_2$  where $u=[u_1,u_2]$ with $\Omega=\mathbb{R}_+^2$,  which induces the following  BD: for $u=[u_1,u_2]$ and $v=[v_1,v_2]$ 
\begin{align}
 \ell_\phi(u,v) =   u_1 \log \frac{u_1}{v_1}+u_2 \log \frac{u_2}{v_2}.  \label{eq:Example_BD_for_Balls}
\end{align} 
The BD in \eqref{eq:Example_BD_for_Balls} is known as generalized I-divergence. 
Fig.~\ref{fig:BregmanBall} compares neighborhoods $\mathcal{B}_{\phi}(r,c)$ and $\tilde{\mathcal{B}}_{\phi}(r,c)$ induced by the BD in \eqref{eq:Example_BD_for_Balls}   to the standard Euclidian ball $\mathcal{B}_{ \|  x \|^2}(r,c)$ where we set $c=(2,2)$ and $r=1$.

\subsection{Variational Characterization of the Bayesian Bregman risk}

With Lemma~\ref{lem:BregmanRemeinder} at our disposal, we are now ready to derive a variational characterization of the Bayesian Bregman risk.

\begin{theorem}\label{thm:BregmanRiskVar} \emph{(Variational Characterization of Bayesian Bregman Risk.)}   
    Let $g: \mathbb{R}^k \to \mathbb{R}^n$. Then, 
    \begin{align}
        &\! \E \left[    \ell_\phi(X, g(Y))   \right]  \notag\\
        &= \sup_{ \psi: \mathbb{R}^n \times \mathbb{R}^k \to \mathbb{R}^n }
       \frac{ \left |\E \left[  (X- g(Y))^T   \psi(X,Y) \right] \right |^2 }{ \E \left[ \| \mathsf{\Delta}_{\phi}^{ -\frac{1}{2}}(X,g(Y))  \psi(X,Y) \|^2 \right]  }  ,  \label{eq:SecondVariationalChar}
    \end{align}
    and equality in \eqref{eq:SecondVariationalChar} is attained if and only if 
    \begin{align} 
        \psi(X,Y) =  \mathsf{\Delta}_{\phi}(X,g(Y))   (X- g(Y)  ). 
    \end{align}
\end{theorem} 

\begin{IEEEproof}
    By using Lemma~\ref{lem:BregmanRemeinder} we have that 
    \begin{align}
        &\! \E \left[    \ell_\phi(X, g(Y))   \right] \\
        &=\E \left[ \| (X- g(Y))^T \mathsf{\Delta}_{\phi}^{\frac{1}{2}}(X,g(Y))  \|^2   \right]\\
        & \ge \frac{ \left | \E \left[  (X- g(Y))^T \mathsf{\Delta}_{\phi}^{\frac{1}{2}}(X,g(Y))   h(X,Y) \right] \right |^2 }{ \E \left[ \| h(X,Y) \|^2 \right]  } ,
    \end{align}
    where the last step follows form the Cauchy--Schwarz inequality for some arbitrary function  $h:\mathbb{R}^n \times \mathbb{R}^k \to \mathbb{R}^n $. 
    Next, we rescale the expression by choosing
    \begin{align}
        h(x,y)=   \mathsf{\Delta}_{\phi}^{-\frac{1}{2}}(x,g(y)) \psi(x,y), 
    \end{align} 
    for some arbitrary function $\psi:\mathbb{R}^n \times \mathbb{R}^k \to \mathbb{R}^n $, which leads to the expression on right side of \eqref{eq:SecondVariationalChar}.  The proof of the equality condition follows by inspection. 
    This concludes the proof. 
\end{IEEEproof}

The variational characterization  in \eqref{eq:SecondVariationalChar} is a generalization of the  Weinstein--Weiss representation of the MSE, which is included as the special case when $\mathsf{\Delta}_{\phi}=\mathsf{I}$ \cite{weinstein1988general}. 

Setting $g(Y) = \E[X|Y]$ yields a variational characterization of the minimum Bayesian risk with respect to a BD, which is an important corollary of the above result.
\begin{corollary}
    \begin{align}
        R_{\phi}( X|Y ) \hspace{-0.05cm} =  \hspace{-0.05cm} \sup_{ \psi: \mathbb{R}^n \times \mathbb{R}^k \to \mathbb{R}^n }
       \frac{ \left |\E \left[  (X- \E[X|Y])^T   \psi(X,Y) \right] \right |^2 }{ \E \left[ \| \mathsf{\Delta}_{\phi}^{ -\frac{1}{2}}(X,\E[X|Y])  \psi(X,Y) \|^2 \right]  }  . 
    \end{align} 
\end{corollary}

\section{Applications}
\label{sec:ApplicationsOfVar}

In this section, we first show a small application of the alternative representation of the BD in Lemma~\ref{lem:BregmanRemeinder}. Second, we present a generalized version of the CR that is specific to a given BD  bound. 

\subsection{Comparing Bregman Risk and the MMSE} 

Our first application shows how Bregman risks can be connected to the ubiquitous case of a risk with a squared error loss.

\begin{prop}\label{thm:MMSEandBregmanRisk} 
    Suppose that $  \kappa_l \mathsf{I} \preceq \mathsf{H}_\phi   \preceq   \kappa_u \mathsf{I} $ for some  constants $ \kappa_l ,  \kappa_u \ge 0$.  Then, 
    \begin{align}
        \kappa_l \mmse(X|Y) \le     R_\phi( X|Y ) \le  \kappa_l \mmse(X|Y). 
    \end{align}
\end{prop}
\begin{IEEEproof}
 The proof follows by using Lemma~\ref{lem:BregmanRemeinder}, that is under the hypothesis of the theorem we have that
 \begin{align}
 \kappa_l \mathsf{I} \preceq    \Delta_\phi(u,v) \preceq \kappa_u \mathsf{I} . 
 \end{align}
 Hence,
 \begin{align}
 \kappa_l \ell_{x^2}(u,v)   \le \ell_\phi(u,v) \le \kappa_u \ell_{x^2}(u,v). 
 \end{align} 
 This concludes the proof. 
\end{IEEEproof}

%

\subsection{A Generalization of the Bayesian CR Bound} 

The classic CR bound allows for lower bounding the MMSE with the Fisher information: for $X\in \Omega \subseteq \mathbb{R}^n$ 
\begin{align}
    \mmse(X|Y) \ge \frac{n^2}{\E \left[ \|  \nabla_X \log f_{YX}(Y,X) \|^2 \right]}, \label{eq:CR-boundMMSE}
\end{align}
where the above holds under the regularity conditions
\begin{subequations}
\begin{align}
    \E \left[   \nabla_X \log f_{YX}(Y,X) |Y =y \right]&=0, \, \forall y; \text{ and } \\
     x f_{YX}(y,x)&=0,\, \forall y, \, \forall x \in \partial \Omega,
\end{align}
where $\partial \Omega$ denotes the boundary of the set $\Omega$ \cite{van2004detection}. The quantity $\E \left[ \|  \nabla_X \log f_{YX}(Y,X) \|^2 \right]$ is known as Fisher information.
\label{eq:RegularityCRbound}
\end{subequations}

The next theorem proposes a generalization of the CR bound.

\begin{theorem}\label{thm:CRboundBayesian} \emph{(Generalized CR-Bound.)}
Suppose that conditions in \eqref{eq:RegularityCRbound} hold. 
Then, 
\begin{align}
\E \left[    \ell_\phi(X, g(Y))   \right]  \hspace{-0.05cm}  \ge \hspace{-0.05cm}   \frac{n^2}{  \E  \hspace{-0.05cm}  \left[ \| \mathsf{\Delta}_{\phi}^{ -\frac{1}{2}}(X,g(Y))  \nabla_X  \log f_{YX}(Y,X) \|^2 \right] }.  \label{eq:CRboundNew}
\end{align}
\end{theorem} 
\begin{IEEEproof}
The proof follows by choosing $\psi(x,y)=\nabla_x \log f_{YX}(y,x)$ in \eqref{eq:SecondVariationalChar}.  Now observe that 
\begin{align}
   &\!\E \left[  (X- g(Y))^T  \nabla_X \log f_{YX}(Y,X) \right] \notag\\
    &=   \E \left[  X^T  \nabla_X \log f_{YX}(Y,X) \right] \notag\\
    &\quad -  \E \left[  g(Y)   \E[\nabla_X \log f_{YX}(Y,X)|Y] \right]\\
    &=\E \left[  X^T  \nabla_X \log f_{YX}(Y,X) \right],
\end{align}
where $\E \left[  g(Y)   \E[\nabla_X \log f_{YX}(Y,X)|Y] \right]=0$  from the assumption in \eqref{eq:RegularityCRbound}. To conclude the proof note that
\begin{align}
   & \E \left[  X^T  \nabla_X  \log f_{YX}(Y,X) \right] \notag\\
    &= \int \int x^T \frac{ \nabla_x f_{YX}(y,x)}{f_{YX}(y,x)}   f_{YX}(y,x) {\rm d}x {\rm d}y\\
    &= \int \int x^T    \nabla_x f_{YX}(y,x)    {\rm d}x {\rm d}y\\
    &=  \sum_{i=1}^n \int \int x_i    \frac{\partial }{ \partial x_i } f_{YX}(y,x)    {\rm d}x {\rm d}y\\ 
    &= -n,
\end{align}
where in the last step we have used integration by parts and that  $ x    f_{YX}(y,x)=0 $ for $x\in \partial \Omega$.  
\end{IEEEproof}

Observe that the quantity in \eqref{eq:CRboundNew} is a generalization of the Fisher information that takes into account the corresponding BD.

\begin{remark} 
    An interesting feature of the CR lower bound in \eqref{eq:CRboundNew} is that it depends on the estimator $g(Y)$.  Note that in the case of the MSE the CR bound in \eqref{eq:CR-boundMMSE} does not depend on the estimator in question and is uniform  over all estimators.   On the one hand, a benefit of this dependence is that one can adapt the lower bound to the estimator in use and potentially get a tighter bound. On the other hand, a drawback might be the computability of such a bound. However, the latter can be addressed by using the CR bound corresponding to $R_\phi( X|Y )$, which is uniform over all estimators, i.e., 
    \begin{align}
        &\! \E \left[    \ell_\phi(X, g(Y))   \right]  \notag\\
        &\ge   R_\phi( X|Y ) \\
        & \ge   \frac{n^2}{  \E \left[ \| \mathsf{\Delta}_{\phi}^{ -\frac{1}{2}}(X,\E[X|Y])  \nabla_X  \log f_{YX}(Y,X) \|^2 \right] }. 
    \end{align} 
\end{remark}

In view of the above discussion, it would be interesting to derive CR lower bounds that hold for an important class  of estimators such as linear estimators.  To that end, we define the linear Bayesian risk as follows:  
\begin{align}
    R_{\phi,\mathsf{L}}( X|Y ) = \E \left[  \ell_{\phi} (  X; g(Y)) \right] \text{ where }  g(Y)= c Y+d.
\end{align} 
We choose not to overload nation for $    R_{\phi,\mathsf{L}}( X|Y ) $ by specifying  coefficients $c$ and $d$ as the values of these will be clear from the context.  Moreover, we will denote by $c^\star$ and $d^\star$  the optimal coefficient under the squared error loss, which in the scalar case  are given by
    \begin{align}
        c^\star&=  \frac{\E\left[(X-\E[X]) (Y-\E[Y]) \right]}{\mathbb{V}(Y)},\\
        d^\star&=\E[X]-  c^\star\E[Y].
    \end{align}

In the rest of the paper, we outline a procedure for applying the CR bound in Theorem~\ref{thm:CRboundBayesian}
 to both  $R_{\phi,\mathsf{L}}( X|Y )$ and $R_{\phi}( X|Y )$. We will consider applying these bounds to two important examples. First, estimation of $X$ in Poisson noise, and second estimation of $X$ in Binomial noise.   Moreover,  for  additional insights, we will also evaluate the CR bounds for the $\mmse(X|Y)$.


\section{Evaluation of  the CR bound for the Poisson Noise Case}
\label{sec:CRboundsEvaluation}

\begin{figure}[tb]  
	\centering   
	\input{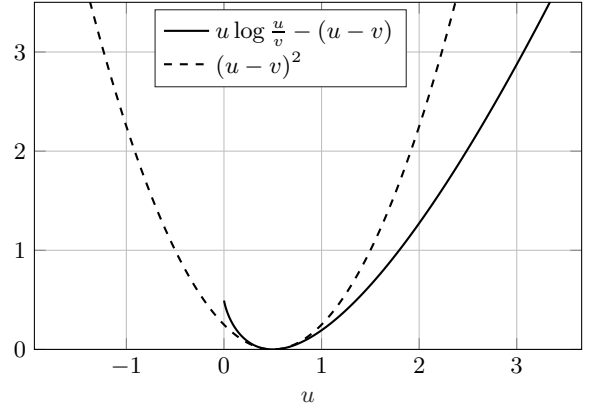}
	\caption{Comparison of the squared error loss to the loss in \eqref{eq:BregmanPoisson} for $v=0.5$.    }
	\label{fig:Comp}
	 \vspace{-0.2cm}
\end{figure}

In this section,  we seek to compute the CR bound in Theorem~\ref{thm:CRboundBayesian}  for a non-trivial case.   Specifically, we consider     $\phi(u)= u \log u$  with $\Omega=\mathbb{R}_{+}$ so that
\begin{align}
 \ell_{\phi} (  u; v)= u\log \frac{u}{v}-(u-v), \,  u,v \in \Omega \label{eq:BregmanPoisson}
\end{align}
which is   natural  for  Poisson noise.   Note that  $\phi''(u)$ is unbounded, and the results of Proposition~\ref{thm:MMSEandBregmanRisk} do not apply. Therefore, it is non-trivial to  compare the Bayesian risk  corresponding to \eqref{eq:BregmanPoisson} and the MMSE.   Fig.~\ref{fig:Comp} compares the squared error loss to the loss in \eqref{eq:BregmanPoisson}.

Next, consider the problem of denoising  a non-negative random  variable in \emph{Poisson noise}. The random Poisson transformation of a non-negative real-valued input random variable $X$ to a non-negative integer-valued output random variable $Y$  will be denoted by
\begin{align}
    Y=  \mathcal{P}(aX), \label{Eq:PoissonTransformationDefinition}
\end{align} 
where  $a>0$ is the \emph{scaling factor}. Concretely, the Poisson noise channel is dictated by the following probability mass function (pmf)
\begin{align}
P_{Y|X}(y|x) = \frac{1}{y!} (ax)^y \eu^{- (ax)}, \label{eq:PoissonTransformation}
\end{align}
where $y = 0, 1, \ldots$ and $x\ge 0$. In words, conditioned on a non-negative input $X$, the output of the Poisson channel is  a non-negative integer-valued random variable $Y$ that is distributed according to  \eqref{eq:PoissonTransformation}.  Note that in \eqref{eq:PoissonTransformation}  we use the convention that  $0^0=1$.   Poisson noise models are  an important family of models with a wide range of applications, including laser communications \cite{gordon1962quantum,shamai1990capacity}.

The first result of this section provides a condition under which the CR bound in \eqref{eq:CRboundNew}  holds.  Note that the output of the Poisson noise channel is discrete.  However, there is no issue in applying  the CR bound in  \eqref{eq:CRboundNew} as differentiability is only required in the $X$ variable, while the $Y$ variable can have an arbitrary support. 

\begin{prop} \emph{(CR Regularity Condition for Poisson Noise.)}
    Let $X \sim f_X$ and $Y=  \mathcal{P}(aX)$. The conditions in \eqref{eq:RegularityCRbound} hold if 
    \begin{align}
        \lim_{x \to 0^{+}}  f_X(x)=0. \label{eq:DensityZeroBehaviour}
    \end{align}
\end{prop}
\begin{IEEEproof}
\begin{align}
&\! \E \left[   \nabla_X \log \left( P_{Y|X}(Y|X) f_X(X) \right) |Y =y \right] \notag\\
&=  \int_0^\infty  \frac{ \nabla_x \left(P_{Y|X}(y|x) f_X(x) \right)}{ P_{Y|X}(y|x) f_X(x)} f_{X|Y}(x|y) {\rm d}x\\
&= \frac{1}{P_Y(y)} \int_0^\infty  \ \nabla_x \left( P_{Y|X}(y|x) f_X(x) \right)  {\rm d}x\\
& = \frac{1}{P_Y(y)}\left(  P_{Y|X}(y|x) f_X(x) \Bigr|_{0}^\infty   \right)\\
& = -\frac{1}{P_Y(y)} \lim_{x \to0^{+}}  P_{Y|X}(y|x) f_X(x)  =0,
\end{align}
where the last step follows from the assumption  in \eqref{eq:DensityZeroBehaviour}.  
This verifies that the CR bound applies. 
\end{IEEEproof}

Before proceeding to apply the CR bound to Bayesian risk with the loss in \eqref{eq:BregmanPoisson}, it is instructive to apply the CR bound to the MMSE.  

\subsection{Estimation with the Squared Loss}

In this section, we evaluate the tightness of the CR bound for the MMSE. 
In order to do this, we also need an upper bounds on the MMSE. Therefore, we begin by finding the best linear estimator for the Poisson noise and the corresponding MSE error, which we will use to upper bound the MMSE.   
\begin{lem}\label{lem:LMMSEpoisson}  \emph{(Best Linear Estimator for the MMSE.)}
    Let $Y=\mathcal{P}(aX)$, $\phi(u)=u^2$, and let
    \begin{align}
        f(c,d)= \E \left[  \ell_{\phi} (  X;c Y+d) \right]. \label{eq:SquareErrorPoisson}
    \end{align} 
    Then,
    \begin{align}
     \mmse(X|Y) &\le \min_{c,d} f(c,d) \label{eq:mmse_lin_bound} \\
     &=  f \left(  \frac{ \mathbb{V}(X)}{a \mathbb{V}(X)+ \E[X]},  \frac{\E^2[X]}{ a \mathbb{V}(X)+ \E[X] }  \right) \label{eq:SOLutiontoLMMSEpoisson}\\
    &=\frac{\mathbb{V}(X) }{ a \frac{\mathbb{V}(X) }{\E[X]}  + 1}.  \label{eq:lmmseBOundPoisson}
    \end{align}
\end{lem}
\begin{IEEEproof}
    The first inequality in \eqref{eq:mmse_lin_bound} follows directly from the definition of the MMSE. It is well-known that the minimizers of \eqref{eq:SquareErrorPoisson} are given by 
    \begin{align}
        c^\star&=  \frac{\E\left[(X-\E[X]) (Y-\E[Y]) \right]}{\mathbb{V}(Y)},\\
        d^\star&=\E[X]-  c^\star\E[Y],
    \end{align}
    and
    \begin{align}
        f(c^\star&,d^\star)= \mathbb{V}(X)-  (c^\star)^2 \mathbb{V}(Y).
    \end{align} 
    The proof is complete by observing that
    \begin{align}
        \E[Y]&=  \E[aX]= a\E[X],\\
    \mathbb{V}(Y)&=  a^2 \mathbb{V}(X)+a\E[X],
    \end{align}
    and 
    \begin{align}
        \E\left[(X-\E[X]) (Y-\E[Y]) \right]= a \mathbb{V}(X). 
    \end{align}
\end{IEEEproof}

Next, we evaluate the CR lower bound for the MMSE. 

\begin{theorem}\label{thm:CR-MMSE} \emph{(CR Bound for the MMSE.)} Let $Y=\mathcal{P}(aX)$ and suppose  \eqref{eq:DensityZeroBehaviour} holds. Then, 
\begin{align}
 \mmse(X|Y)  \ge   \frac{1}{a   \E \left[ \frac{1}{X} \right] +\E \left[ \rho_X^2(X)  \right]}.  \label{eq:CR-MMSE}
\end{align} 
\end{theorem}
\begin{IEEEproof}
Observe that the score function is readily computed to be 
\begin{align}
    \nabla_x \log \left( P_{Y|X}(y|x) f_X(x) \right)&= \frac{  \nabla_x P_{Y|X}(y|x)}{ P_{Y|X}(y|x)} + \frac{\nabla_x f_X(x)}{f_X(x)}\\
    &= \frac{y}{x}-a+\rho_X(x).  \label{eq:ScoreFunction_Poisson}
\end{align}
Therefore,  the denominator in the CR bound is given by 
\begin{align}
&\E \left[ \left(  \frac{Y}{X}-a+\rho_X(X) \right)^2 \right] \notag\\
&= \E \left[ \left(  \frac{Y}{X}-a \right)^2 \right] +  2 \E \left[  \left(  \frac{Y}{X}-a\right )\rho_X(X)  \right]+\E \left[ \rho_X^2(X)  \right]\\
&= \E \left[ \left(  \frac{Y}{X}-a \right)^2 \right] +\E \left[ \rho_X^2(X)  \right] \label{eq:CR-MMSEcomp:eq1}\\
&= a   \E \left[ \frac{1}{X} \right] +\E \left[ \rho_X^2(X)  \right] , \label{eq:CR-MMSEcomp:eq2}
\end{align}
where in \eqref{eq:CR-MMSEcomp:eq1} we have used that 
\begin{align}
\E \left[ \left(\frac{Y}{X}-a \right) \Big| X \right]= \frac{  \E \left[ \left(Y-aX \right) | X \right]}{X} =0; 
\end{align} 
and in \eqref{eq:CR-MMSEcomp:eq2} we have used  the variance of the Poisson distribution, so that 
\begin{align}
\E \left[ \left( Y-aX \right)^2 |X  \right]=aX. 
\end{align} 
This concludes the proof. 
\end{IEEEproof}

  From Theorem~\ref{thm:CR-MMSE} and the upper bound in \eqref{eq:lmmseBOundPoisson} we conclude that under the aforementioned regularity conditions  
  \begin{align}
  \mmse(X|Y)= \Theta \left( \frac{1}{a} \right). 
  \end{align}
  In other words, the CR bound is an effective  lower bound for large values of $a$.

We now proceed to evaluate the bounds in Lemma~\ref{lem:LMMSEpoisson} and  Theorem~\ref{thm:CR-MMSE}  for the case when $X$ is distributed according to a gamma distribution,  whose pdf is given by 
\begin{align}
f(x)= \frac{\alpha^\theta}{ \Gamma(\theta)  }  x^{\theta-1} \eu^{-\alpha x},\, x \ge 0, \label{eq:pdfGamma}
\end{align}
 where $\theta>0$ is the shape parameter and $\alpha>0$ is the rate parameter.  We denote the distribution with the pdf in \eqref{eq:pdfGamma} by $\mathsf{Gam}(\alpha, \theta)$.  The choice of a gamma distribution is dictated by the fact that it is  a conjugate prior of the Poisson distribution \cite{diaconis1979conjugate}.  The next lemma compiles several properties of the gamma distribution needed in this section and the next.

\begin{lem}\label{lem:FactsAboutGamma} 
    Suppose that  $X \sim \mathsf{Gam}(\alpha, \theta)$ and that $Y=\mathcal{P}(aX)$. Then,
    \begin{align}
        \E[X ]&=\frac{\theta}{\alpha}, \\
        \mathbb{V}(X)&=\frac{\theta}{\alpha^2},\\
        \E \left[ \frac{1}{X^n} \right]&= \left \{ \begin{array}{cc} 
        \infty,  & \theta \le n \\
        \alpha^n \frac{ \Gamma(\theta-n)}{ \Gamma(\theta)  }  , & \theta>n 
        \end{array} \right. ,  \label{eq:FractionalMOmentGamma}  \\
        \E[ \rho_X^2(X) ]&=  \left \{ \begin{array}{cc}  
        \infty, & \theta \le 2  \\
        \frac{\alpha^2 }{\theta-2} & \theta > 2\end{array}  \right. , \label{Eq:FisherInformation}  \\
        \E[ \rho_X^2(X)X ]&=\left \{ \begin{array}{cc} 
        \infty, & \theta \le 1 \\
        \alpha, & \theta > 1
        \end{array}  \right.  , \\
        \E[X|Y=y]&=   \frac{1}{\alpha+a}y +\frac{\theta}{\alpha+a}. \label{eq:EstimatorForGamma}
    \end{align}
\end{lem}

\begin{IEEEproof} 
  See Appendix~\ref{app:lem:FactsAboutGamma}.
\end{IEEEproof}

The next result evaluates the MMSE and the CR lower bound for the case when $X \sim \mathsf{Gam}(\alpha, \theta)$. 
 
\begin{prop} \emph{(Gamma Prior.)}\label{prop:GammaMMSE} 
    Suppose that  $X \sim \mathsf{Gam}(\alpha, \theta)$ and $Y~=~\mathcal{P}(aX)$. Then,  the following  statements hold:
    \begin{itemize}[leftmargin=*]
        \item  The upper bound in \eqref{eq:lmmseBOundPoisson} is tight, that is
            \begin{align}
                \mmse(X|Y)=\frac{\mathbb{V}(X) }{ a \frac{\mathbb{V}(X) }{\E[X]}  + 1}  = \frac{\theta}{\alpha (a+\alpha)}, \label{eq:ExactMMSEgamma}
            \end{align} 
            with $\mathbb{V}(X)=\frac{\theta}{\alpha^2}$  and $\E(X)=\frac{\theta}{\alpha}$;
        \item The CR regularity condition in \eqref{eq:DensityZeroBehaviour} holds for $\theta>1$.  Moreover, the bound  in     \eqref{eq:CR-MMSE} reduces to 
            \begin{align}
                \mmse(X|Y) \ge  \left \{  \begin{array}{cc} 
                    0,   & 1 <  \theta  \le 2\\
                    \frac{\theta-1}{ \alpha  \left(a+ \alpha \frac{\theta-1}{\theta-2} \right)  }, &  \theta \ge 2
                \end{array}
                \right.  .  \label{eq:CR-MMSE-Gamma}
            \end{align} 
    \end{itemize} 
\end{prop} 
\begin{IEEEproof}
    Observe that the optimal MMSE estimator for $X \sim \mathsf{Gam}(\alpha, \theta)$ is given by \eqref{eq:EstimatorForGamma}, while the estimator that achieves the upper bound in \eqref{eq:lmmseBOundPoisson} is given in \eqref{eq:SOLutiontoLMMSEpoisson}  by 
    \begin{align}
        \frac{ \mathbb{V}(X)}{a \mathbb{V}(X)+ \E[X]} y+ \frac{\E^2[X]}{ a \mathbb{V}(X)+ \E[X] }  &=  \frac{1}{a+\alpha}y +  \frac{ \theta }{\theta+\alpha},
    \end{align}
    where we have used that $\E[X ]=\frac{\theta}{\alpha}$ and $\mathbb{V}(X)=\frac{\theta}{\alpha^2}$. 
    Since the two estimators agree, the upper bound is achieved with equality. 
    
    The fact that the regularity condition in \eqref{eq:DensityZeroBehaviour} holds for $\theta>1$ follows from the limit 
    \begin{align}
        \lim_{x \to 0^{+}} x^{\theta-1} \eu^{-\alpha x} = \left \{  \begin{array}{cc} 
        \infty   &   \theta  <1\\
        1 , &  \theta =1 \\
        0, &  \theta>1
        \end{array} \right. .
    \end{align} 
    The proof of \eqref{eq:CR-MMSE-Gamma} now follows by  inserting \eqref{eq:FractionalMOmentGamma} and \eqref{Eq:FisherInformation} into \eqref{eq:CR-MMSE}. 
 \end{IEEEproof}
  
\begin{figure}[tb]  
	\centering   
%
%
\pgfplotsset{every axis plot/.append style={thick}}
\begin{tikzpicture}
\small
\begin{axis}[%
width=\columnwidth,
height=0.7\columnwidth,
xmin=0,
xmax=15,
xlabel style={font=\color{white!15!black}},
xlabel={$a$},
ymin=0,
ymax=0.7,
axis background/.style={fill=white},
xmajorgrids,
ymajorgrids,
legend style={legend cell align=left, align=left, draw=white!15!black}
]
\addplot [color=black]
  table[row sep=crcr]{%
0	0.680272108843537\\
0.151515151515152	0.634493366660258\\
0.303030303030303	0.594487479733381\\
0.454545454545455	0.559227249618709\\
0.606060606060606	0.527915533514638\\
0.757575757575758	0.499924253900924\\
0.909090909090909	0.474751834268451\\
1.06060606060606	0.451992877687988\\
1.21212121212121	0.4313161678212\\
1.36363636363636	0.412448443944507\\
1.51515151515152	0.395162256017243\\
1.66666666666667	0.379266750948167\\
1.81818181818182	0.364600596619158\\
1.96969696969697	0.351026486543985\\
2.12121212121212	0.338426828017639\\
2.27272727272727	0.326700326700327\\
2.42424242424242	0.315759257487322\\
2.57575757575758	0.305527265993889\\
2.72727272727273	0.295937584073177\\
2.87878787878788	0.286931571167725\\
3.03030303030303	0.278457514133828\\
3.18181818181818	0.270469633636587\\
3.33333333333333	0.262927256792287\\
3.48484848484848	0.255794124486474\\
3.63636363636364	0.249037808467285\\
3.78787878787879	0.242629218439821\\
3.93939393939394	0.236542183356032\\
4.09090909090909	0.230753094189218\\
4.24242424242424	0.225240597911405\\
4.39393939393939	0.219985334311046\\
4.54545454545455	0.214969708813758\\
4.6969696969697	0.210177695688173\\
4.84848484848485	0.205594666998941\\
5	0.201207243460765\\
5.15151515151515	0.19700316399021\\
5.3030303030303	0.192971171276534\\
5.45454545454545	0.189100911122572\\
5.60606060606061	0.185382843660469\\
5.75757575757576	0.181808164839403\\
5.90909090909091	0.178368736825036\\
6.06060606060606	0.175057026152459\\
6.21212121212121	0.1718660486433\\
6.36363636363636	0.168789320239374\\
6.51515151515152	0.165820813024471\\
6.66666666666667	0.162954915806627\\
6.81818181818182	0.160186398718509\\
6.96969696969697	0.157510381366045\\
7.12121212121212	0.154922304117178\\
7.27272727272727	0.152417902175419\\
7.42424242424242	0.149993182128085\\
7.57575757575758	0.147644400697955\\
7.72727272727273	0.145368045460552\\
7.87878787878788	0.143160817318121\\
8.03030303030303	0.141019614546387\\
8.18181818181818	0.138941518251863\\
8.33333333333333	0.136923779096303\\
8.48484848484848	0.134963805161343\\
8.63636363636364	0.133059150840692\\
8.78787878787879	0.131207506659775\\
8.93939393939394	0.129406689933728\\
9.09090909090909	0.127654636184287\\
9.24242424242424	0.125949391244609\\
9.39393939393939	0.12428910398855\\
9.54545454545454	0.122672019627523\\
9.6969696969697	0.121096473523907\\
9.84848484848485	0.119560885475164\\
10	0.118063754427391\\
10.1515151515152	0.116603653581146\\
10.3030303030303	0.115179225855991\\
10.4545454545455	0.113789179683459\\
10.6060606060606	0.112432285101019\\
10.7575757575758	0.111107370122218\\
10.9090909090909	0.109813317360487\\
11.0606060606061	0.108549060886155\\
11.2121212121212	0.107313583298104\\
11.3636363636364	0.106105912993151\\
11.5151515151515	0.104925121617755\\
11.6666666666667	0.103770321687997\\
11.8181818181818	0.102640664365028\\
11.969696969697	0.101535337374235\\
12.1212121212121	0.100453563057441\\
12.2727272727273	0.0993945965482967\\
12.4242424242424	0.098357724061876\\
12.5757575757576	0.0973422612902274\\
12.7272727272727	0.096347551896295\\
12.8787878787879	0.0953729660992457\\
13.030303030303	0.094417899344797\\
13.1818181818182	0.0934817710546443\\
13.3333333333333	0.0925640234495526\\
13.4848484848485	0.0916641204410989\\
13.6363636363636	0.0907815465874391\\
13.7878787878788	0.0899158061088254\\
13.9393939393939	0.0890664219589215\\
14.0909090909091	0.0882329349482634\\
14.2424242424242	0.087414902916479\\
14.3939393939394	0.0866118999501325\\
14.5454545454545	0.0858235156432863\\
14.6969696969697	0.0850493543980825\\
14.8484848484848	0.0842890347628413\\
15	0.0835421888053467\\
};
\addlegendentry{MMSE in \eqref{eq:ExactMMSEgamma}}

\addplot [color=black,dashed]
  table[row sep=crcr]{%
0	0.226757369614512\\
0.151515151515152	0.218861918026263\\
0.303030303030303	0.211497788886753\\
0.454545454545455	0.204613095238095\\
0.606060606060606	0.19816249324446\\
0.757575757575758	0.192106182326231\\
0.909090909090909	0.186409083206236\\
1.06060606060606	0.181040157998683\\
1.21212121212121	0.175971844504879\\
1.36363636363636	0.171179582944289\\
1.51515151515152	0.166641417966975\\
1.66666666666667	0.162337662337662\\
1.81818181818182	0.158250611422817\\
1.96969696969697	0.154364299747404\\
2.12121212121212	0.150664292562663\\
2.27272727272727	0.147137506688068\\
2.42424242424242	0.1437720559404\\
2.57575757575758	0.140557117301303\\
2.72727272727273	0.137482814648169\\
2.87878787878788	0.13454011741683\\
3.03030303030303	0.131720752005748\\
3.18181818181818	0.129017124091016\\
3.33333333333333	0.126422250316056\\
3.48484848484848	0.12392969806219\\
3.63636363636364	0.121533532206386\\
3.78787878787879	0.119228267938435\\
3.93939393939394	0.117008828847995\\
4.09090909090909	0.114870509607352\\
4.24242424242424	0.112808942672546\\
4.39393939393939	0.110820068506951\\
4.54545454545455	0.108900108900109\\
4.6969696969697	0.107045543012845\\
4.84848484848485	0.105253085829107\\
5	0.10351966873706\\
5.15151515151515	0.101842421997963\\
5.3030303030303	0.100218658892128\\
5.45454545454545	0.0986458613577258\\
5.60606060606061	0.0971216669609747\\
5.75757575757576	0.0956438570559082\\
5.90909090909091	0.0942103460089072\\
6.06060606060606	0.0928191713779428\\
6.21212121212121	0.0914684849492766\\
6.36363636363636	0.0901565445455291\\
6.51515151515152	0.0888817065287653\\
6.66666666666667	0.0876424189307625\\
6.81818181818182	0.0864372151500864\\
6.96969696969697	0.085264708162158\\
7.12121212121212	0.084123585194249\\
7.27272727272727	0.0830126028224285\\
7.42424242424242	0.0819305824519589\\
7.57575757575758	0.0808764061466068\\
7.72727272727273	0.079849012775842\\
7.87878787878788	0.0788473944520106\\
8.03030303030303	0.0778705932323375\\
8.18181818181818	0.0769176980630725\\
8.33333333333333	0.0759878419452887\\
8.48484848484848	0.0750801993038018\\
8.63636363636364	0.0741939835424255\\
8.78787878787879	0.0733284447703486\\
8.93939393939394	0.0724828676858197\\
9.09090909090909	0.071656569604586\\
9.24242424242424	0.0708488986216669\\
9.39393939393939	0.0700592318960576\\
9.54545454545454	0.0692869740488788\\
9.6969696969697	0.0685315556663136\\
9.84848484848485	0.0677924318994207\\
10	0.0670690811535882\\
10.1515151515152	0.0663610038610039\\
10.3030303030303	0.0656677213300698\\
10.4545454545455	0.064988774666194\\
10.6060606060606	0.0643237237588445\\
10.7575757575758	0.063672146330169\\
10.9090909090909	0.0630336370408573\\
11.0606060606061	0.0624078066492681\\
11.2121212121212	0.0617942812201562\\
11.3636363636364	0.0611927013796173\\
11.5151515151515	0.0606027216131343\\
11.6666666666667	0.0600240096038415\\
11.8181818181818	0.0594562456083455\\
11.969696969697	0.0588991218676376\\
12.1212121212121	0.0583523420508196\\
12.2727272727273	0.057815620729528\\
12.4242424242424	0.0572886828810999\\
12.5757575757576	0.0567712634186623\\
12.7272727272727	0.056263106746458\\
12.8787878787879	0.0557639663388421\\
13.030303030303	0.0552736043414904\\
13.1818181818182	0.0547917911934648\\
13.3333333333333	0.0543183052688756\\
13.4848484848485	0.0538529325369627\\
13.6363636363636	0.0533954662395029\\
13.7878787878788	0.0529457065845206\\
13.9393939393939	0.0525034604553482\\
14.0909090909091	0.0520685411341475\\
14.2424242424242	0.0516407680390592\\
14.3939393939394	0.0512199664742038\\
14.5454545454545	0.0508059673918064\\
14.6969696969697	0.0503986071657656\\
14.8484848484848	0.0499977273760284\\
15	0.0496031746031746\\
};
\addlegendentry{CR bound in \eqref{eq:CR-MMSE-Gamma}}

\end{axis}

\end{tikzpicture}%
%
	\caption{Evaluation of the  exact MMSE  in \eqref{eq:ExactMMSEgamma} and the CR bound in \eqref{eq:CR-MMSE-Gamma}. The parameters are set to $\alpha=2.1$ and $\theta=3$.}
	\label{fig:MMSEPoissonCr}
		 \vspace{-0.2cm}
\end{figure}
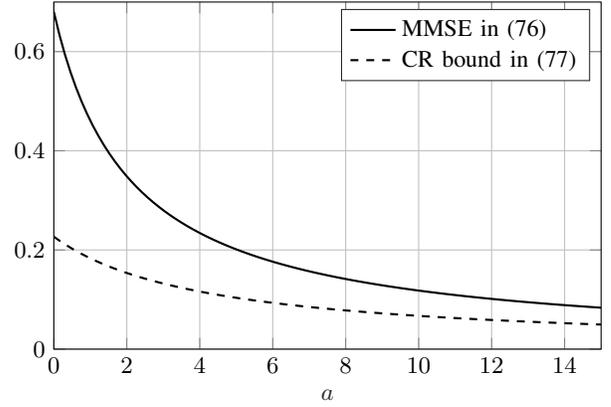

  Fig.~\ref{fig:MMSEPoissonCr} compares the exact MMSE and the CR lower bound evaluated in Proposition~\ref{prop:GammaMMSE}.

\begin{remark}  The fact that the bound in \eqref{eq:lmmseBOundPoisson} is attained by a gamma distribution suggest that a gamma distribution is \emph{the  least favorable prior} distribution for the MMSE under the mean and variance constraint. 
To put it differently, for $Y=\mathcal{P}(aX)$ the maximizer of 
\begin{align}
  P_{X^\star}=\arg \max_{P_X: \E[X]=\mu, \mathbb{V}(X)=P} \mmse(X|Y), \label{eq:LFPMMSEpoissonChannel}
\end{align} 
is given by $ P_{X^\star}= \mathsf{Gam}(\alpha^\star, \theta^\star)$ with $\alpha^\star= \left( \frac{\mu}{P} \right)^{ \frac{1}{4}}$ and $\theta^\star=\mu^{ \frac{3}{4}} P^{ \frac{1}{4}}$.  We also note that the maximizer in \eqref{eq:LFPMMSEpoissonChannel} is unique; see \cite[Remark~9]{dytso2019estimationPoisson} for the details. 
\end{remark}

\subsection{Estimation with the Natural Bregman Divergence} 

Before evaluating the CR bound in Theorem~\ref{thm:CRboundBayesian} for  $\phi(u)= u \log u$,  we present two ancillary lemmas.  The first result provides bounds on $\Delta_\phi(u,v)$. 
  \begin{lem}\label{lem:BoundRatioTerm}  Let  $\phi(u)= u \log u$.  Then, 
  \begin{align}
   u \le     \frac{1}{ \Delta_\phi(u,v)} \le  \frac{4u}{3} + \frac{2v}{3}.  \label{eq:BoundOnTheTilt}
  \end{align}
  \end{lem}
  \begin{IEEEproof}
We first show the upper bound. To that end, let $T$ be a random variable on $[0,1]$ with a pdf given by $f_T(t)=2 (1-t)$. Then,
\begin{align}
    \frac{1}{ \Delta_\phi(u,v)}&=\frac{1}{ \int_0^1  \frac{(1-t)}{(1-t)u + tv}  {\rm d}  t } \\
    &= \frac{1}{ \frac{1}{2}\E \left[  \frac{1}{(1-T)u + Tv} \right] } \\
        &\le 2  \E \left[(1-T)u + Tv \right] \label{eq:JensensStep} \\
    &= 2\E \left[1-T\right] u + 2\E\left[T \right] v \\
    &= \frac{4u}{3} + \frac{2v}{3}, 
\end{align}
where in \eqref{eq:JensensStep} we have used Jensen's inequality.
The proof of the lower bound follows since  $v \ge 0$ and
\begin{align}
    \int_0^1  \frac{(1-t)}{(1-t)u + tv}  {\rm d}  t \le \frac{1}{u}. 
\end{align}
This concludes the proof. 
  \end{IEEEproof}

We now proceed to evaluate the CR lower bound in Theorem~\ref{thm:CRboundBayesian}  for  linear estimators.     Note, that  in the previous section, it  was not possible to derive  estimator-depend bounds using the CR bound in \eqref{eq:CRboundNew} as the term  $ \Delta_\phi(u,v)$ is constant for $\phi(u)=u^2$.   However, as shown next, for $ \phi(u)= u \log u$ the bound will depend on the estimator.  
\begin{theorem} \emph{(CR Bound for Linear Estimators.)} \label{thm:CRboundPoisson}
    Let $\phi(u)= u \log u$  and $Y=\mathcal{P}(aX)$ and   suppose  \eqref{eq:DensityZeroBehaviour} holds.  Then, for the linear estimator $g(y)=c_1y+c_2$, we have that 
    \begin{align}
    R_{\phi,\mathsf{L}}( X|Y )   \ge  \frac{1}{D_{c_1,c_2}}, \label{eq:CR-BregmanPoisson}
    \end{align}
    where
    \begin{align}
       D_{c_1,c_2}&= \frac{4}{3} \left(  a    + \E \left[  \rho_X^2(X) X    \right] \right) \notag\\
        &\quad +\frac{2c_1}{3} \left(a^2+a \E \left[ \frac{1}{X} \right] +a\E \left[   \rho_X^2(X)   X  \right] \right) \notag\\
        &\quad + \frac{2c_2}{3} \left(a   \E \left[ \frac{1}{X} \right] +\E \left[ \rho_X^2(X)  \right]\right).
    \end{align}
\end{theorem}

\begin{IEEEproof}
    See Appendix~\ref{app:thm:CRboundPoisson}.  
\end{IEEEproof}

Note that we always have an inequality  $R_{\phi}( X|Y )  \le   R_{\phi,\mathsf{L}}( X|Y )$. Therefore, it is interesting  to note that the steps leading to the proof of Theorem~\ref{thm:CRboundPoisson}  can also be used to  lower bound the performance of   $R_{\phi}( X|Y ) $.     This is possible in view of the fact that for a  large set of distributions the optimal estimator $\E[X|Y]$ can be bounded by a linear function \cite[Theorem~10]{dytso2019estimationPoisson}. 
\begin{lem}\label{lem:BoundOnPoissonCE} 
    Let $Y=\mathcal{P}(aX)$ and suppose that 
    \begin{align}
\sup_{n  \ge 0} \left|  \frac{ \mathcal{L}_{X}^{(n+1)}(a)  }{ (n+1) \mathcal{L}_{X}^{(n)}(a)  } \right|=c<\infty,\label{eq:LimitSupermumCondition}
\end{align} 
where $\mathcal{L}_{X}(t)=\E\left[\eu^{-tX} \right], t\ge 0$ and $\mathcal{L}_{X}^{(n)}(t)$   is the Laplace transform of $X$ and  its  $n$-th derivative.
    Then,   
    \begin{align}
       \E[X|Y=y]\le c(y+1), \quad y=0,1,\ldots  \label{eq:LinearityBound}
    \end{align}
\end{lem}

 All of the distributions used in this paper  will satisfy the condition in \eqref{eq:LimitSupermumCondition}.   
 
Using  Lemma~\ref{lem:FactsAboutGamma} and Theorem~\ref{thm:CRboundPoisson} we have the following universal CR bound. 
\begin{theorem} \emph{(Universal CR Bound.)}  
 Let $\phi(u)= u \log u$  and $Y=\mathcal{P}(aX)$ and   suppose  \eqref{eq:DensityZeroBehaviour} holds.   Then, 
  \begin{align}
    R_{\phi}( X|Y )   \ge  \frac{1}{D_{c,c}},
    \end{align}
    where $c$ is defined in \eqref{eq:LimitSupermumCondition}. 
\end{theorem} 
\begin{IEEEproof}
Using the CR bound in \eqref{eq:CRboundNew}  and the expression for the score function in \eqref{eq:ScoreFunction_Poisson} we have that 
\begin{align}
   &\! \E \left[ \frac{(  \frac{\rm d}{{\rm d}X} \log f_{YX}(Y,X) )^2}{\Delta_\phi(X,\E[X|Y])} \right]\\
   &= \E \left[ \frac{\left(  \frac{Y}{X}-a+\rho_X(X) \right)^2}{\Delta_\phi(X,\E[X|Y])} \right]\\
    &\le \E \left[ \left(  \frac{Y}{X}-a+\rho_X(X) \right)^2\left( \frac{4X}{3} + \frac{2\E[X|Y]}{3} \right) \right] \label{eq:BoundOnTiltedFisher}\\
    &\le \E \left[ \left(  \frac{Y}{X}-a+\rho_X(X) \right)^2\left( \frac{4X}{3} + \frac{2 c(Y+1)}{3} \right) \right] \label{eq:apply_linear_bound_poisson}\\
    &=D_{c,c} , \label{eq:final_step_CR_universal_poisson}
\end{align}
where \eqref{eq:BoundOnTiltedFisher} follows by using the bound in  \eqref{eq:BoundOnTheTilt};   \eqref{eq:apply_linear_bound_poisson} follows by using the bound in \eqref{eq:LinearityBound}; and \eqref{eq:final_step_CR_universal_poisson} follows by using the same steps as in the proof of Theorem~\ref{thm:CRboundPoisson}.  This concludes the proof. 
\end{IEEEproof}

%
 
In order to evaluate the tightness of the CR bound, we also provide an exact expression for   $R_\phi( X|Y )$, which is amenable to numerical computations.   

\begin{prop} \emph{(Gamma Prior.)} \label{prop:EvaluationOfbregmanForGamma} 
    Let $\phi(u)= u \log u$,  $Y=\mathcal{P}(aX)$ and $X \sim \mathsf{Gam}(\alpha, \theta)$. Then,  the following statements hold:
    \begin{itemize}[leftmargin=*]
    \item  (Linear MMSE Estimator is Optimal) Let $g(y)= c^\star y+d^\star$. Then, 
    \begin{align}
      R_\phi( X|Y ) =R_{\phi,\mathsf{L}}( X|Y ) .   \label{eq:LinearAndOptimalRisk}
    \end{align} 
    \item  (CR  Bound)  The CR regularity condition holds for $\theta>1$.  Moreover, for $g(y)= c^\star y+d^\star$ the bounds  in \eqref{eq:CR-BregmanPoisson} reduce to  
    \begin{align}
        D_{c^\star, d^\star}&= \frac{4}{3} \left(  a    + \alpha \right) +\frac{2}{3}   \frac{1}{\alpha+a} \left(a^2+a \frac{\alpha}{\theta-1} +a \alpha \right) \notag\\
        &\quad + \frac{2}{3} \frac{\theta}{\alpha+a} \left(a   \frac{\alpha}{\theta-1} + \frac{\alpha^2 }{\theta-2} \right), \label{eq:CRevaluatedForGamma}
        \end{align}
            for $\theta >2$ and $   D_{c^\star, d^\star}=\infty$ for $\theta \in [1,2]$. 
        \item (Exact Expression) 
           \begin{align}
        R_\phi( X|Y ) =  \E[X \log X ]  -B, \label{eq:BregmanDiverGenceGamma}
    \end{align} 
    where
          \begin{align}
        \E[X \log X ] =\frac{\theta \left(\log\left(\frac{1}{\alpha}\right)+\psi \left(\theta+1\right)\right)}{\alpha},
    \end{align} 
    where $\psi $ is the digamma function, and
    \begin{align}
       \hspace{-0.22cm} B&=\E  \hspace{-0.02cm} \left[  \hspace{-0.02cm}  \left ( \frac{Y}{\alpha+a} +\frac{\theta}{\alpha+a} \right)  \hspace{-0.05cm}  \log \left ( \frac{Y}{\alpha+a} +\frac{\theta}{\alpha+a} \right) \hspace{-0.02cm} \right] , \label{Eq:ExpressionForTheBterm}
    \end{align}
    and where $Y$  has a negative binomial distribution with the  pmf given by
    \begin{align}
        P_{Y}(y)=   \frac{ a^y  \alpha^\theta}{  \left(\alpha+ a\right)^{\theta+y} } \binom{\theta+y-1}{y},\, y=0,1,\ldots \label{eq:NegativeBinomial}
    \end{align}
  \item   (Alternative Upper and Lower Bounds) The expression in \eqref{Eq:ExpressionForTheBterm} can be further upper and lower bounded as follows: 
    \begin{align}
       \hspace{-0.25cm} & \E \left[   X   \log \left ( \frac{aX}{\alpha+a} +\frac{\theta}{\alpha+a} \right) \right]  \ge B  \notag\\
        & \ge  \E \left[   \left ( \frac{ aX}{\alpha+a} +\frac{\theta}{\alpha+a} \right)   \log \left ( \frac{aX}{\alpha+a} +\frac{\theta}{\alpha+a} \right) \right] . \label{eq:BoundsOnB}
    \end{align}

    \end{itemize}

\end{prop} 

\begin{IEEEproof}
    See Appendix~\ref{app:prop:EvaluationOfbregmanForGamma}. 
\end{IEEEproof}

\begin{figure}[tb]
	\centering   
	\input{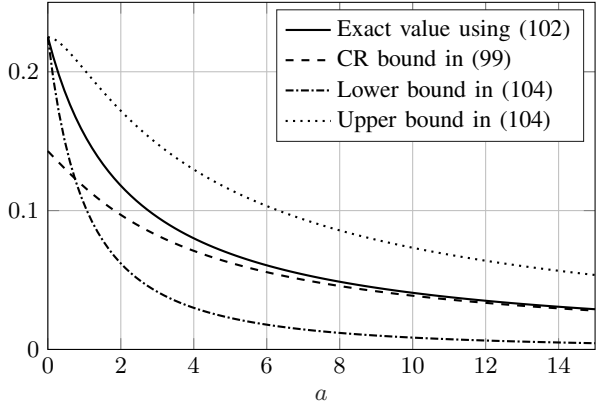}%
	\caption{Evaluation of the  exact Bregman risk in \eqref{Eq:ExpressionForTheBterm}, the CR bound in \eqref{eq:CRevaluatedForGamma}, and the bounds in \eqref{eq:BoundsOnB}. The parameters are set to $\alpha=2.1$ and $\theta=3$.}
	\label{fig:BregmanPoissonGammaInput}
		 \vspace{-0.2cm}
\end{figure}

Fig.~\ref{fig:BregmanPoissonGammaInput} compares the CR bound in \eqref{eq:CRevaluatedForGamma} to the exact value and bounds on  $R_\phi( X|Y )$ computed in Proposition~\ref{prop:EvaluationOfbregmanForGamma}. 

It is also instructive to loosen the lower  bound in \eqref{eq:BoundsOnB} to
\begin{align}
B \ge  \E \left[   \left ( \frac{ aX}{\alpha+a}  \right)   \log \left ( \frac{aX}{\alpha+a}  \right) \right],
\end{align}
which implies that 
\begin{align} 
R_\phi( X|Y ) \hspace{-0.05cm} \le  \hspace{-0.05cm}   \frac{ \E \left[X \log  X \right]}{\alpha+a}  -  \frac{ (a  \E \left[   X \right] +\theta) \log \left ( \frac{a}{\alpha+a}  \right)}{\alpha+a}  . 
\end{align} 
Therefore,  the new CR bound is of the same order as the upper bound and 
\begin{align} 
R_\phi( X|Y ) =\Theta \left( \frac{1}{a} \right). 
\end{align}
This conclusion demonstrates that the new CR bound is effective. 
Note, that both $R_\phi( X|Y ) $ and the MMSE, as shown in \eqref{eq:ExactMMSEgamma},  have the same order.  

\begin{figure}[tb]
	\centering   
	\input{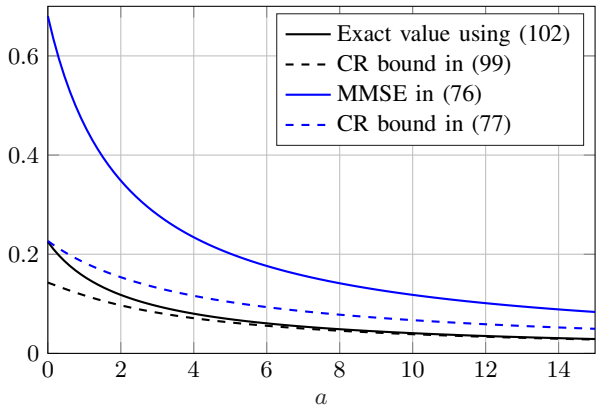}%
	\caption{Comparison of the MMSE and the Bregman risk with $\phi(u)=u \log u$. The parameters are set to $\alpha=2.1$ and $\theta=3$.     }
	\label{fig:BregmanvsMMSE}
		 \vspace{-0.2cm}
\end{figure}

Fig.~\ref{fig:BregmanvsMMSE} concludes this section by comparing the MMSE in \eqref{eq:ExactMMSEgamma}, the CR bound in \eqref{eq:CR-MMSE-Gamma} specific to the MMSE, the Bayesian Bregman risk in \eqref{eq:BregmanDiverGenceGamma} and the CR bound in \eqref{eq:CRevaluatedForGamma}. 

\section{Evaluation of  the CR bound for the Binomial Noise Case} 
\label{sec:BinamialSection}

In this section, we consider estimation of a random variable $X \in (0,1)$ in Binomial noise.  Specifically, the random binomial transformation of an input random variable $X \in (0,1)$ to a non-negative integer-valued output random variable $Y \in \{0,1, \ldots, n \}$  where $n$ is some fixed positive integer will be denoted by
\begin{align}
    Y=  \mathcal{B}_n(aX), \label{Eq:BinomialTransformationDefinition}
\end{align} 
where  $a \in [0,1]$ is the \emph{scaling factor}. Furthermore, the conditional distribution of the Binomial noise model is given by  the following pmf: for   $y =0,1, \ldots, n$
\begin{align}
P_{Y|X}(y|x) = {{n} \choose {y} }   (ax )^y (1-ax)^{n-y}.
\end{align}
The Bregman divergence natural for the Binomial noise is generated by the function $ \phi(u)=u \log \frac{u}{1-u}$ where $\Omega=[0,1]$ and is given by 
\begin{align}
\ell_\phi(u,v)= u \log \frac{u(1-v)}{v(1-u)} - \frac{(u - v)}{1-v}, \,    u,v \in \Omega.  \label{eq:BregmanBinomial}
\end{align} 
Fig. 2 compares the squared error loss to the loss in \eqref{eq:BregmanBinomial}.
  
  \begin{figure}[tb]  
	\centering   
	\input{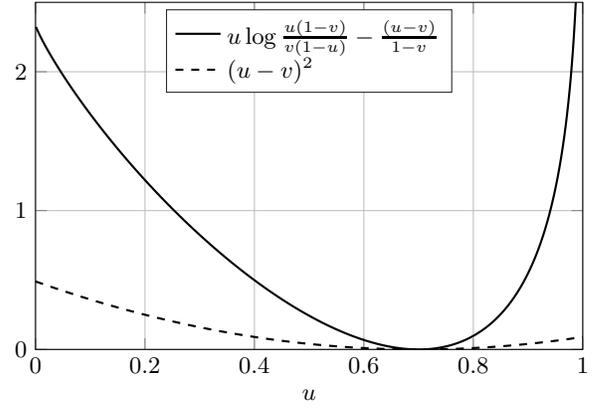}
	\caption{Comparison of the squared error loss to the loss in \eqref{eq:BregmanBinomial} for $v=0.7$.    }
	\label{fig:CompBregmanMSEBInomial}
		 \vspace{-0.2cm}
\end{figure}%

The first result of this section provides a condition under which the CR bound in \eqref{eq:CRboundNew}  holds for the Binomial model.

\begin{prop} \emph{(CR Regularity Condition for Binomial Noise.)}
    Let $X \sim f_X$ and $Y=  \mathcal{B}_n(aX)$. The conditions in \eqref{eq:RegularityCRbound} hold if 
    \begin{align}
        \lim_{x \to 0^{+}}  f_X(x)=0, \text{ and }    \lim_{x \to 1^{-}}  f_X(x)=0. \label{eq:AssumptionForBinomial}
    \end{align}
\end{prop}
\begin{IEEEproof}
\begin{align}
&\! \E \left[   \nabla_X \log \left( P_{Y|X}(Y|X) f_X(X) \right) |Y =y \right] \notag\\
&=  \int_0^\infty  \frac{ \nabla_x \left(P_{Y|X}(y|x) f_X(x) \right)}{ P_{Y|X}(y|x) f_X(x)} f_{X|Y}(x|y) {\rm d}x\\
&= \frac{1}{P_Y(y)} \int_0^1  \ \nabla_x \left( P_{Y|X}(y|x) f_X(x) \right)  {\rm d}x\\
& = \frac{1}{P_Y(y)}\left(  P_{Y|X}(y|x) f_X(x) \Bigr|_{0}^1   \right)\\
& = \frac{   \lim_{x \to1^{-}}  P_{Y|X}(y|x) f_X(x) -  \lim_{x \to0^{+}}  P_{Y|X}(y|x) f_X(x)}{P_Y(y)} \\
&=0,
\end{align}
where the last step follows from the assumption  in \eqref{eq:AssumptionForBinomial}.  
This verifies that the CR bound applies. 
\end{IEEEproof}

Similarly  to the Poisson example,  we begin by applying the CR bound to the MMSE. 

\subsection{Estimation with the Squared Loss}

We begin by deriving the best linear estimator and the associated MSE.  
\begin{lem}\label{lem:LMMSEbinomial}   \emph{(Best Linear Estimator for the MMSE.)}
    Let $Y=\mathcal{B}_n(aX)$, $\phi(u)=u^2$, and let
    \begin{align}
        f(c,d)= \E \left[  \ell_{\phi} (  X;c Y+d) \right]. \label{eq:SquareErrorBinomial}
    \end{align} 
    Then,
    \begin{align}
    \mmse(X|Y)
     &\le \min_{c,d} f(c,d) \label{eq:mmse_lin_bound_binomial} \\
     &=  f \left(  c^\star ,  d^\star  \right)  \label{eq:OptimalCoeffiCint_lmmse_binomial}\\
     &= \frac{ \mathbb{V}(X)}{  1 + \frac{ na  \mathbb{V}(X)}{ \E[ X(1-aX)]}},  \label{eq:lmmse_binomial}
    \end{align}
    where $ c^\star= \frac{an \mathbb{V}(X)}{   \E[ na X(1-aX)] +  (na)^2 \mathbb{V}(X)}$ and  $d^\star=   \frac{  \E[ na X(1-aX)]  \E[X]}{   \E[ na X(1-aX)] +  (na)^2 \mathbb{V}(X)}$.
\end{lem}
\begin{IEEEproof} 
We use a proof similar to that in Lemma~\ref{lem:LMMSEpoisson}. 

First, note that 
\begin{align}
\E[Y]&=\E[ \E[Y|X] ]= na\E[X], \\
\E[Y^2]&=  \E[\E[Y^2|X]] \notag\\
&=  \E[ na X(1-aX)] + (na)^2 \E[  X^2 ]. 
\end{align}
Hence,
\begin{align}
\mathbb{V}(Y)=\E[ na X(1-aX)]  + (na)^2 \mathbb{V}(X),
\end{align}
and
\begin{align}
&\E[(X-\E[X])(Y-\E[Y])]\notag\\
& = \E[(X-\E[X])\E[(Y-\E[Y])|X]] =an \mathbb{V}(X). 
\end{align}

Second, the optimal coefficients for the linear estimator are given by 
\begin{align}
  c^\star&=  \frac{\E\left[(X-\E[X]) (Y-\E[Y]) \right]}{\mathbb{V}(Y)}\\
  &=   \frac{an \mathbb{V}(X)}{   \E[ na X(1-aX)] +  (na)^2 \mathbb{V}(X)},     \\
        d^\star&=\E[X]-  c^\star\E[Y]\\
        &=\E[X]-   \frac{ (an)^2 \mathbb{V}(X) \E[X]}{   \E[ na X(1-aX)] +  (na)^2 \mathbb{V}(X)}\\
        &=   \frac{  \E[ na X(1-aX)]  \E[X]}{   \E[ na X(1-aX)] +  (na)^2 \mathbb{V}(X)},
\end{align}
and  the best error due to the linear estimator is given by 
\begin{align}
f(c^\star,d^\star)&= \mathbb{V}(X)-  (c^\star)^2 \mathbb{V}(Y)\\
&= \mathbb{V}(X)-\frac{ (an)^2 \mathbb{V}^2(X)}{   \E[ na X(1-aX)]  + (na)^2 \mathbb{V}(X)}\\
&= \frac{  \E[ na X(1-aX)]  \mathbb{V}(X)}{   \E[ na X(1-aX)]  + (na)^2 \mathbb{V}(X)} .
\end{align} 
This concludes the proof. 
\end{IEEEproof}

The CR bound for the MMSE is computed next.

\begin{theorem}\label{thm:CR-MMSE-Binomial} \emph{(CR Bound for the MMSE.)} Let $Y=\mathcal{B}_n(aX)$ and suppose  \eqref{eq:AssumptionForBinomial} holds. Then, 
\begin{align}
 \mmse(X|Y)  \ge   \frac{1}{n a \E \left[  \frac{ 1 }{ X (1-aX) }  \right]  +  \E[\rho_X^2(X)  ]}.  \label{eq:CR-MMSE_Binomial}
\end{align} 
\end{theorem}
\begin{IEEEproof}
Observe that the score function is given by
\begin{align}
    \nabla_x \log \left( P_{Y|X}(y|x) f_X(x) \right)&= \frac{  \nabla_x P_{Y|X}(y|x)}{ P_{Y|X}(y|x)} + \frac{\nabla_x f_X(x)}{f_X(x)}\\
    &= \frac{y}{x}- \frac{a (n-y)}{1-ax}+\rho_X(x)\\
    &= \frac{an x-y}{ax^2-x}+\rho_X(x). \label{eq:ScoreFunctionBinomial+Input}
\end{align}
Hence,  the denominator in the CR bound in \eqref{eq:CR-boundMMSE} is given by 
\begin{align}
 &\E \left[ \left(   \frac{an X-Y}{aX^2-X}+\rho_X(X) \right)^2 \right] \notag\\
&=  \E \left[ \left(   \frac{an X-Y}{aX^2-X} \right)^2 \right]  +  \E[\rho_X^2(X)  ]  \notag\\
&  +2 \E \left[ \left(   \frac{an X-Y}{aX^2-X} \right)\rho_X(X) \right] \\
&=  \E \left[ \left(  \frac{an X-Y}{aX^2-X} \right)^2 \right]  +  \E[\rho_X^2(X)  ] \label{eq:CrossScoreFunctionTerm} \\
&=  \E \left[  \frac{ \E[ \left(   an X-Y\right)^2| X ]}{  \left(  aX^2-X \right)^2}  \right]  +  \E[\rho_X^2(X)  ]\\
&=  \E \left[  \frac{ n a X (1-aX)}{  \left(  aX^2-X \right)^2}  \right]  +  \E[\rho_X^2(X)  ] \label{Eq:VarianceOfBinomial}\\
&=   n a \E \left[  \frac{ 1 }{ X (1-aX) }  \right]  +  \E[\rho_X^2(X)  ], \label{eq:TermCR_Binomail_mmse}
\end{align}
where \eqref{eq:CrossScoreFunctionTerm} follows by  using 
\begin{align}
 \E \left[   an X-Y  |X \right]=0;
\end{align} 
and \eqref{Eq:VarianceOfBinomial} follows by using 
\begin{align}
 \E \left[ \left( an X-Y \right)^2 | X \right] = n a X (1-aX).
\end{align} 
This concludes the proof.  
\end{IEEEproof} 

Observe that the upper bound  in \eqref{eq:lmmse_binomial} and the lower bound in \eqref{eq:CR-MMSE_Binomial}  have the same rate of convergence as $n$ goes to infinity, which implies that 
\begin{align}
\mmse(X|Y)= \Theta \left(\frac{1}{n} \right ). 
\end{align}
This illustrates that the CR bound is order tight.  

To further evaluate the performance of the bounds we will use the beta distribution as prior on $X$.  The pdf of the beta distribution is give by 
\begin{align}
f(x) &= c_{\alpha,\beta} x^{\alpha-1} (1-x)^{\beta-1}, \, \alpha>0, \beta>0, \label{eq:pdf_Beta}\\
 c_{\alpha,\beta} &= \left \{ \begin{array}{cc} \frac{\Gamma(\alpha+\beta)}{\Gamma(\alpha) \Gamma(\beta)} ,  & \alpha>0, \beta>0,\\
0 ,& \text{ otherwise}.   
 \end{array} \right.  \label{eq:Normalization_Constant_Beta}
\end{align}
We denote the distribution with the pdf in \eqref{eq:pdf_Beta} by $\mathsf{Beta}(\alpha,\beta)$. 

The next lemma compiles several properties of the beta distribution needed in this section and the next.

\begin{lem} \label{lem:Beta_Quantities}  
  Suppose that  $X \sim \mathsf{Beta}(\alpha,\beta)$ and that $Y=\mathcal{B}_n(aX)$. Then, the moments and the variance are given by 
\begin{align}
&\mathbb{V}(X)=\frac{\alpha \beta}{ (\alpha+\beta)^2 (\alpha+\beta+1)}, \label{eq:VarianceBeta}\\
& \E \left[  \hspace{-0.02cm} \frac{ X^k }{ (1-aX)^m }  \right] \notag\\
 &=   \frac{c_{\alpha,\beta}  F_{2,1} (  \alpha+k,m; \alpha+ \beta+k;a)}{c_{\alpha+k,\beta}}  ,  k,m \in \mathbb{R}, \label{eq:Term_HPfunction}
  \end{align}
where $F_{1,2}$ is a hypergeometric function. The score function and the related moments are given by
\begin{align}
\rho_X(x)&=\frac{\alpha-1}{x}-\frac{1-\beta}{1-x}, \, x\in (0,1) ,\\
\E[\rho_X^2(X) X^k ] &= (\alpha-1)^2   \frac{c_{\alpha,\beta}}{c_{\alpha-2+k,\beta}}+ (1-\beta)^2   \frac{c_{\alpha,\beta}}{c_{\alpha+k,\beta-2}} \notag\\
&+2 (1-\alpha)(1-\beta)   \frac{c_{\alpha,\beta}}{c_{\alpha+k-1,\beta-1}} , k \in \mathbb{R},
\end{align} 
and the conditional expectation is given by 
\begin{align}
&\E[ X|Y=y] \notag\\
&=\frac{ y  \mathbb{V}(X)  +   \E[  X ] \E[  X (1-aX)  ]}{  na   \mathbb{V}(X)  +  \E[ X (1-aX) ] },  \,y=0,1,\ldots, n.   \label{eq:CE_Beta_Distribution}
\end{align} 
\end{lem} 
\begin{IEEEproof}
See Appendix~\ref{app:lem:Beta_Quantities}.
\end{IEEEproof} 

The next result evaluates the upper bound in \eqref{eq:lmmse_binomial} and lower bound in \eqref{eq:CR-MMSE_Binomial} for the case when $X$ is according to the beta distribution. 

\begin{prop}\label{prop:EvaluationForBetaMMSE}  
\emph{(Beta Prior.)}  Suppose that $X \sim \mathsf{Beta}(\alpha,\beta)$ and $Y=\mathcal{B}_n(aX)$. Then, the following statements hold:
\begin{itemize}[leftmargin=*]
\item The upper bound in \eqref{eq:lmmse_binomial} is tight, that is
\begin{align}
\mmse(X|Y)=  \frac{ \mathbb{V}(X)}{  1 + \frac{ na  \mathbb{V}(X)}{  \E[X] - a \E[X^2] }} , \label{eq:MMSE_Beta_Binomial}
\end{align} 
where  $\E[X]=  \frac{\alpha}{\alpha+\beta}$,     $\E[X^2]=\frac{(\alpha+2)(\alpha+1)}{(\alpha+\beta+2) (\alpha+\beta+1) }$ and  $\mathbb{V}(X)=\frac{\alpha \beta}{ (\alpha+\beta)^2 (\alpha+\beta+1)}$.
\item  The CR regularity condition in \eqref{eq:AssumptionForBinomial} hols if $\alpha>1 $ and $\beta>1$.   Moreover, the bound  in     \eqref{eq:CR-MMSE_Binomial} reduces to 
            \begin{align}
                \mmse(X|Y) \ge  \left \{  \begin{array}{cc} 
                     \frac{1}{n a \E \left[  \frac{ 1 }{ X}  \right]  +  \E[\rho_X^2(X)  ]} ,    &   \beta >2,  \alpha >2 \\
                     0,  & \text{ otherwise} 
                \end{array}
                \right.  ,  \label{eq:CR-MMSE-Binomial_Beta}
            \end{align} 
\end{itemize}
where    $ \E \left[ \frac{1}{X} \right] $ and  $\E[\rho_X^2(X)]$ are given in Lemma~\ref{lem:Beta_Quantities}. 
\end{prop}
\begin{IEEEproof} 
Observe that the optimal MMSE estimator for $X \sim \mathsf{Beta}(\alpha, \beta)$ is given by \eqref{eq:CE_Beta_Distribution}, while the estimator that achieves the upper bound in \eqref{eq:mmse_lin_bound_binomial} is given in \eqref{eq:OptimalCoeffiCint_lmmse_binomial}  by 
    \begin{align}
 c^\star  y+ d^\star=    \frac{ y \mathbb{V}(X)   +  \E[ X(1-aX)]  \E[X]}{  na \mathbb{V}(X)+   \E[ X(1-aX)] } .
    \end{align} 
   Since the two estimators agree, the upper bound is achieved with equality. 

The fact that the regularity condition in \eqref{eq:AssumptionForBinomial} holds for  $\alpha>1$ and $\beta>1$ is a consequence of the following limits:
\begin{align}
 \lim_{x \to 0^{+} }  x^{\alpha-1} (1-x)^{\beta-1}&=\left \{ \begin{array}{cc}    0 &     \alpha>1 \\
 1 & \alpha=1\\
 \infty &  \alpha<1
   \end{array}  \right.  , \\
    \lim_{x \to 1^{-} }  x^{\alpha-1} (1-x)^{\beta-1}&=\left \{ \begin{array}{cc}    0 &     \beta>1 \\
 1 & \beta=1\\
 \infty &  \beta<1
   \end{array}  \right. .
\end{align} 
The rest of the proofs follows by using the properties  in  Lemma~\ref{lem:Beta_Quantities}.
\end{IEEEproof}

\begin{figure}[tb]
	\centering   
%
%
\pgfplotsset{every axis plot/.append style={thick}}
\begin{tikzpicture}
\small
\begin{axis}[%
width=\columnwidth,
height=0.7\columnwidth,
xmin=0,
xmax=100,
xlabel style={font=\color{white!15!black}},
xlabel={$n$},
ymin=0,
ymax=0.035,
axis background/.style={fill=white},
xmajorgrids,
ymajorgrids,
legend style={legend cell align=left, align=left, draw=white!15!black}
]
\addplot [color=black]
  table[row sep=crcr]{%
1	0.0334458562306664\\
2	0.0297783240213944\\
3	0.0268356403409931\\
4	0.0244222427584906\\
5	0.0224071138579504\\
6	0.0206991822490446\\
7	0.0192331768172807\\
8	0.017961094573257\\
9	0.01684684434986\\
10	0.0158627677693068\\
11	0.0149873122431761\\
12	0.0142034342237742\\
13	0.0134974786362333\\
14	0.0128583766621383\\
15	0.0122770611279967\\
16	0.0117460336131488\\
17	0.0112590392447585\\
18	0.0108108191708767\\
19	0.0103969198938889\\
20	0.0100135447876921\\
21	0.00965743729746496\\
22	0.00932578820587978\\
23	0.00901616137252031\\
24	0.0087264337911486\\
25	0.0084547468444381\\
26	0.00819946638951923\\
27	0.00795914986259364\\
28	0.00773251900357809\\
29	0.00751843711160792\\
30	0.00731588997696907\\
31	0.00712396981434364\\
32	0.00694186166031653\\
33	0.00676883180518159\\
34	0.00660421791273645\\
35	0.00644742054753266\\
36	0.006297895881095\\
37	0.00615514939005463\\
38	0.0060187303923067\\
39	0.00588822729399733\\
40	0.00576326344174012\\
41	0.00564349349201857\\
42	0.00552860022407045\\
43	0.00541829173431457\\
44	0.00531229896007378\\
45	0.00521037348836741\\
46	0.00511228561220717\\
47	0.0050178226023835\\
48	0.00492678716737579\\
49	0.00483899607792099\\
50	0.00475427893606175\\
51	0.00467247707127265\\
52	0.00459344254861804\\
53	0.00451703727589695\\
54	0.00444313219843829\\
55	0.00437160657166917\\
56	0.00430234730283139\\
57	0.004235248354297\\
58	0.00417021020186144\\
59	0.00410713934219375\\
60	0.00404594784431725\\
61	0.00398655294059608\\
62	0.00392887665322661\\
63	0.00387284545268937\\
64	0.00381838994501573\\
65	0.00376544458507252\\
66	0.00371394741337377\\
67	0.00366383981419772\\
68	0.00361506629302362\\
69	0.0035675742715118\\
70	0.00352131389843453\\
71	0.00347623787512846\\
72	0.00343230129418382\\
73	0.00338946149021384\\
74	0.00334767790166208\\
75	0.00330691194270669\\
76	0.00326712688441125\\
77	0.00322828774435286\\
78	0.00319036118403021\\
79	0.00315331541341927\\
80	0.00311712010210241\\
81	0.00308174629644856\\
82	0.00304716634236941\\
83	0.00301335381321819\\
84	0.00298028344243629\\
85	0.00294793106058673\\
86	0.00291627353644476\\
87	0.00288528872184393\\
88	0.0028549554000011\\
89	0.00282525323706724\\
90	0.00279616273667127\\
91	0.00276766519724366\\
92	0.00273974267192326\\
93	0.00271237793086696\\
94	0.00268555442579561\\
95	0.00265925625662317\\
96	0.00263346814002742\\
97	0.00260817537983186\\
98	0.00258336383907824\\
99	0.00255901991367792\\
100	0.00253513050753929\\
};
\addlegendentry{MMSE in \eqref{eq:MMSE_Beta_Binomial}}

\addplot [color=black,dashed]
  table[row sep=crcr]{%
1	0.00882705073044494\\
2	0.00859651240952141\\
3	0.00837770964363638\\
4	0.0081697685768299\\
5	0.0079719000109259\\
6	0.00778338939606711\\
7	0.00760358820859967\\
8	0.00743190649703214\\
9	0.00726780641552746\\
10	0.00711079659558889\\
11	0.00696042723186508\\
12	0.00681628577855494\\
13	0.00667799316969492\\
14	0.00654520049040644\\
15	0.00641758603755574\\
16	0.0062948527176946\\
17	0.0061767257379771\\
18	0.00606295055227598\\
19	0.00595329103018842\\
20	0.00584752782121301\\
21	0.00574545689025038\\
22	0.005646888203853\\
23	0.00555164454942568\\
24	0.00545956047194035\\
25	0.00537048131474333\\
26	0.00528426235275712\\
27	0.0052007680078578\\
28	0.00511987113748009\\
29	0.0050414523885992\\
30	0.00496539961018571\\
31	0.00489160731805046\\
32	0.00481997620670932\\
33	0.00475041270351756\\
34	0.00468282856086418\\
35	0.00461714048268905\\
36	0.0045532697819993\\
37	0.00449114206642399\\
38	0.00443068694916536\\
39	0.00437183778298511\\
40	0.00431453141511236\\
41	0.00425870796117825\\
42	0.00420431059647602\\
43	0.00415128536301705\\
44	0.00409958099100557\\
45	0.0040491487334905\\
46	0.00399994221307348\\
47	0.0039519172796598\\
48	0.00390503187833516\\
49	0.00385924592653717\\
50	0.0038145211997676\\
51	0.0037708212251605\\
52	0.00372811118228327\\
53	0.0036863578106036\\
54	0.00364552932310567\\
55	0.00360559532558384\\
56	0.00356652674118342\\
57	0.00352829573979464\\
58	0.00349087567193957\\
59	0.00345424100682202\\
60	0.00341836727423765\\
61	0.00338323101006656\\
62	0.00334880970509338\\
63	0.00331508175692004\\
64	0.0032820264247555\\
65	0.0032496237868836\\
66	0.0032178547006257\\
67	0.00318670076462909\\
68	0.00315614428332506\\
69	0.00312616823341231\\
70	0.00309675623223243\\
71	0.00306789250791392\\
72	0.00303956187117043\\
73	0.00301174968864732\\
74	0.00298444185771821\\
75	0.00295762478264039\\
76	0.00293128535198423\\
77	0.00290541091725812\\
78	0.00287998927265543\\
79	0.00285500863585558\\
80	0.00283045762981569\\
81	0.00280632526549365\\
82	0.00278260092544758\\
83	0.00275927434826017\\
84	0.00273633561373994\\
85	0.00271377512885462\\
86	0.0026915836143546\\
87	0.00266975209204747\\
88	0.0026482718726868\\
89	0.00262713454444098\\
90	0.00260633196190984\\
91	0.00258585623565906\\
92	0.00256569972224398\\
93	0.00254585501469646\\
94	0.00252631493344978\\
95	0.00250707251767827\\
96	0.0024881210170298\\
97	0.00246945388373036\\
98	0.00245106476504142\\
99	0.00243294749605187\\
100	0.00241509609278716\\
};
\addlegendentry{CR bound in \eqref{eq:CR-MMSE-Binomial_Beta}}

\end{axis}

\end{tikzpicture}%
%
	\caption{Evaluation of the exact MMSE  in \eqref{eq:MMSE_Beta_Binomial} and the CR bound in \eqref{eq:CR-MMSE-Binomial_Beta}. The parameters are set to $a=0.8$ and  $\alpha=3, \beta=2.5$.   }
	\label{fig:MMSE_CR_Binomial}
		 \vspace{-0.2cm}
\end{figure}
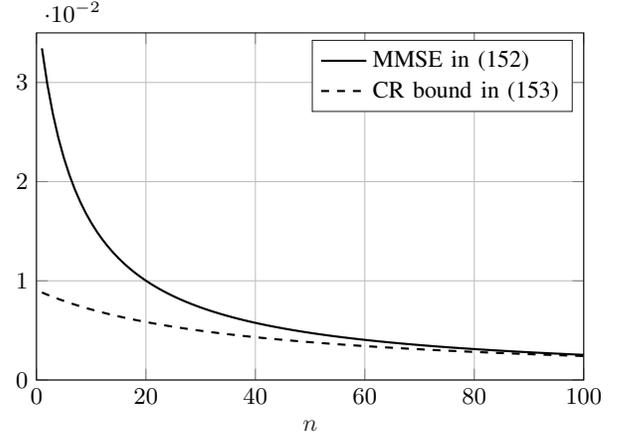%

\subsection{Estimation with the Natural BD}
In this section, we apply the CR bound to the Bayesian risk with BD in \eqref{eq:BregmanBinomial}.   Similarly to the Poisson case, we first provide  a polynomial bound on  $\Delta_\phi^{-1}(u,v)$. 

\begin{lem} 
Let $\phi=u \log \frac{u}{1-u}$ with $u \in \Omega$.  Then, 
\begin{align}
    \frac{1}{ \Delta_\phi(u,v)} 
        &  \le \frac{4u}{3} + \frac{2v}{3} - u^2- \frac{ 2uv}{3}-\frac{v^2}{3} \label{eq:SeconBoundOnDelta}\\
        & \le \frac{4u}{3} + \frac{2v}{3} - u^2  \label{eq:ThirdBoundOnDelta}. 
    \end{align} 
\end{lem} 
\begin{IEEEproof}
First, note that $\phi''(u)=   \frac{1}{u(1-u)^2}$. Second, similarly to the proof of Lemma~\ref{lem:BoundRatioTerm}, let $T$ be a random variable on $[0,1]$ with a pdf given by $f_T(t)=2 (1-t)$, and let $T_{u,v}=(1-T)u + Tv$.  Then,
\begin{align}
    \frac{1}{ \Delta_\phi(u,v)}&=\frac{1}{ \int_0^1  (1-t)  \phi''((1-t)u + tv)    {\rm d}  t } \\
    &= \frac{1}{ \frac{1}{2}\E \left[  \frac{1}{T_{u,v}(1-T_{u,v})^2} \right] } \\
        &\le 2  \E \left[ T_{u,v}(1-T_{u,v})^2 \right],  \label{eq:BoundOnDelta_Binamial_Jensen}
        \end{align}
        where \eqref{eq:BoundOnDelta_Binamial_Jensen} follows from Jensen's inequality. Furthermore, 
        \begin{align}
         \E \left[ T_{u,v}(1-T_{u,v})^2 \right]  &\le   \E \left[ T_{u,v}(1-T_{u,v})\right]\\
         & = \frac{2u}{3} + \frac{v}{3} - \frac{u^2}{2}- \frac{ uv}{3}-\frac{v^2}{6} \label{eq:EvaluationTuv}\\
         & \le   \frac{2u}{3} + \frac{v}{3} - \frac{u^2}{2}. 
        \end{align} 
      where in \eqref{eq:EvaluationTuv} we have used that
        \begin{align}
        \E \left[ T_{u,v}\right]&=\frac{2u}{3} + \frac{v}{3},\\
          \E \left[ T_{u,v}^2\right]&= \frac{u^2}{2}+ \frac{ uv}{3}+\frac{v^2}{6} .
        \end{align} 
        This concludes the proof.  
                 \end{IEEEproof}
            
   We now proceed to evaluate the CR lower bound in Theorem~\ref{thm:CRboundBayesian} for linear estimators. 
            \begin{theorem}\label{thm:CR_for_Linear_Estimators_Binomial} \emph{(CR Bound for Linear Estimators.)} Let $Y=\mathcal{B}_n(aX)$ and suppose  \eqref{eq:AssumptionForBinomial} holds. Then, for the linear estimator $g(y)=c_1y+c_2$, we have that 
            \begin{align}
             R_{\phi,\mathsf{L}}( X|Y )  \ge   \frac{1}{  \sum_{0 \le r \le 2, 0 \le m \le 2}   \kappa_{r,m}  G_X(a;r,m) }, \label{eq:CR_Binomial_Linear}
            \end{align}
            where 
            \begin{align}
            \kappa_{0,2}&=\frac{c_1^2 }{3},   &\kappa_{0,1}&= \frac{2c_1(1-c_2) }{3}, &\kappa_{1,1}&=\frac{2c_1 }{3}, \notag\\
             \kappa_{2,0}&=-1,  &\kappa_{1,0}&=\frac{2(2-c_2)}{3},   &\kappa_{0,0}&= \frac{2c_2-c_2^2}{3},
            \end{align} 
            and $k_{r,m}=0$ otherwise, and the function  $G_X $ is defined as 
            \begin{align}
            G_X(a;r,m) &= \sum_{k=0}^{m+2}   d_{k,m,n} a^k    \E \left[ \frac {   X^{k+r-2} }{  \left( 1 -a X \right)^2}  \right] \notag\\
&+  \sum_{k=0}^m  s_{m,k}  (n)_k  a^k  \E[\rho_X^2(X) X^{k+r} ] \notag\\
&+ 2\sum_{k=0}^{m+1}  h_{k,m,n} a^{k+1}   \E \left[   \frac{ X^{k+r-1} }{( 1- aX )^2}  \right] \notag\\
           &+  2 \sum_{k=0}^{m+1}  h_{k,m,n} a^{k} (k+r-1) \E \left[   \frac{ X^{k+r-2} }{1-aX}  \right], 
            \end{align} 
            and where $(n)_k$ is the falling factorial,  $s_{m,k}$ the Stirling's number of the second kind  and
            \begin{align}
            d_{k,m,n}&=  n^2   (n)_{k-2} s_{m,k-2}  - 2 n   (n)_{k-1} s_{m+1,k-1}+ (n)_k s_{m+2,k}, \\
             h_{k,m,n} & =  n (n)_{k-1}   s_{m,k-1}    -    (n)_k s_{m+1,k}. 
            \end{align} 
            \end{theorem}
            \begin{IEEEproof}
         See   Appendix~\ref{app:thm:CR_for_Linear_Estimators_Binomial}.
                \end{IEEEproof} 
                While the bound in \eqref{eq:CR_Binomial_Linear}  consists of long algebraic expression, in principle it is not difficult to compute provided that the terms of the type $   \E \left[ \frac {   X^{m} }{  \left( 1 -a X \right)^k} \right] $ have a closed-form expression.  For example, as shown in Lemma~\ref{lem:Beta_Quantities}, such terms can be compute for the beta distribution.  
        
We now proceed to show  a universe CR bound that holds for all estimators. 
             \begin{theorem} 
             \emph{(Universal CR Bound.)} For $Y=\mathcal{B}_n(aX)$ and 
       suppose  \eqref{eq:AssumptionForBinomial} holds. Then,
        \begin{align}
         R_\phi( X|Y ) 
         & \ge    \frac{1}{  \sum_{ k=0}^2 c_k  \left(\E \left[  \frac{n a X^k }{ X (1-aX) }   \right]  +  \E[\rho_X^2(X) X^k  ]  \right)   }, \label{eq:CR_Universal_Bound_Binomial}
        \end{align}
        where $c_0=\frac{2}{3}, c_1=\frac{4}{3}$ and $c_2=-1$.        
        \end{theorem}
        \begin{IEEEproof}
   The proof of the lower bound follows by using the CR bound
\begin{align}
   &\! \E \left[ \frac{(  \frac{\rm d}{{\rm d}X} \log f_{YX}(Y,X) )^2}{\Delta_\phi(X,\E[X|Y])} \right]\\
   &= \E \left[ \frac{\left(  \frac{an X-Y}{aX^2-X}+\rho_X(X) \right)^2}{\Delta_\phi(X,\E[X|Y])} \right]\\
    &\le \E \left[ \left(   \frac{an X-Y}{aX^2-X}+\rho_X(X) \right)^2 \hspace{-0.04cm}\left(  \frac{4X}{3} + \frac{2\E[X|Y]}{3} -X^2 \right) \right] \label{eq:simplification_1}\\
      &\le \E \left[ \left(   \frac{an X-Y}{aX^2-X}+\rho_X(X) \right)^2\left(  \frac{4X}{3} + \frac{2}{3} -X^2 \right) \right],\label{eq:simplification_2}
\end{align}
where in \eqref{eq:simplification_1} we have used the bound in \eqref{eq:ThirdBoundOnDelta}; and  in \eqref{eq:simplification_2} we have used the bound $\E[X|Y] \le 1$. 

The terms in  \eqref{eq:simplification_2} are of the form 
\begin{align}
& \E \left[ \left(   \frac{an X-Y}{aX^2-X}+\rho_X(X) \right)^2 X^k  \right] \notag\\
 &=  \E \left[ \left(   \frac{an X-Y}{aX^2-X} \right)^2 X^k \right]  +  \E[\rho_X^2(X) X^k  ]   \notag\\
& +2 \E \left[ \left(   \frac{an X-Y}{aX^2-X} \right)\rho_X(X) X^k \right] \\
 &=  \E \left[ \left(   \frac{an X-Y}{aX^2-X} \right)^2 X^k \right]  +  \E[\rho_X^2(X) X^k  ] \label{eq:E[ an X-Y | X]=0} \\
  &=  \E \left[  \frac{\E[(an X-Y)^2|X]}{ (   aX^2-X )^2}  X^k \right]  +  \E[\rho_X^2(X) X^k  ] \\
    &=  \E \left[  \frac{n a X (1-aX)}{ (   aX^2-X )^2}  X^k \right]  +  \E[\rho_X^2(X) X^k  ]  \label{eq: n a X (1-aX)}\\
     &=  \E \left[  \frac{n a X^k }{ X (1-aX) }   \right]  +  \E[\rho_X^2(X) X^k  ],  \label{eq:Bound_first_Term_CR}
 \end{align}
 where in \eqref{eq:E[ an X-Y | X]=0} we have used that 
 \begin{align}
 \E[ an X-Y | X]=0;
 \end{align} 
and where  in  \eqref{eq: n a X (1-aX)} we have used that 
\begin{align}
 \E \left[ \left( an X-Y \right)^2 | X \right] = n a X (1-aX).
\end{align} 
 
Combining  \eqref{eq:simplification_2} and \eqref{eq:Bound_first_Term_CR}  concludes the proof.
\end{IEEEproof}

To evaluate the tightness of the CR bound we compute it for the beta prior. Moreover, we provide an exact expression for $R_{\phi}(X|Y)$ which  is suitable for numerical computation. 
\begin{prop}\label{prop:Evaluation_Beta_Bregman} 
\emph{(Beta Prior.)}  Let $Y=\mathcal{B}_n(aX)$ and $X\sim \mathsf{Beta}(\alpha,\beta)$.  Then, the following statements hold: 
\begin{itemize}[leftmargin=*]
   \item  (Linear MMSE Estimator is Optimal) Let $g(y)= c^\star y+d^\star$. Then, 
    \begin{align}
      R_\phi( X|Y ) =R_{\phi,\mathsf{L}}( X|Y ) .   \label{eq:LinearAndOptimalRiskBinomial}
    \end{align} 
    \item (CR bound)  The CR regularity condition in \eqref{eq:AssumptionForBinomial} hols if $\alpha>1 $ and $\beta>1$.  Moreover, the terms in the function   $ G_X(a;r,m)$ in the CR bound in Theorem~\ref{thm:CR_for_Linear_Estimators_Binomial} are computed in Lemma~\ref{lem:Beta_Quantities}. 
    \item (An Exact Expression) 
    \begin{align}
R_{\phi}(X|Y)=  \E\left[X \log  \frac{X}{1-X}\right]   + B,  \label{Eq:ExpressionForTheBtermBinaomial}
\end{align}  
where 
\begin{align}
 \E\left[X \log  \frac{X}{1-X}\right]&=  \frac{\alpha}{\alpha+\beta} ( \psi(\alpha+1)-\psi(\beta) ) ,  \\
B&= \E \left[ ( c^\star Y+d^\star) \log \frac{1-c^\star Y-d^\star }{ c^\star Y+d^\star }  \right],
\end{align}
where $Y$ has the following pmf: for  $y=0,\ldots, n$ 
\begin{align}
P_Y(y) 
&= \frac{c_{\alpha,\beta}   {{n} \choose {y} }  a^y     F_{2,1}( \alpha+y,y-n;\beta+\alpha+y;a )  }{  c_{\alpha+y,\beta} }  .
\end{align}
\item (An Upper Bound)  The expression in \eqref{Eq:ExpressionForTheBtermBinaomial} can be further upper bounded as follows: 
\begin{align}
B \le   \E \left[ ( c^\star anX +d^\star) \log \frac{1-c^\star  anX-d^\star }{ c^\star  anX+d^\star }   \right].  \label{eq:UpperBoundBinoamialBeta}
\end{align}
\end{itemize}

\end{prop} 
\begin{IEEEproof}
See Appendix~\ref{app:prop:Evaluation_Beta_Bregman}. 
\end{IEEEproof} 

\begin{figure}[tb]
	\centering   
%
%
\definecolor{mycolor1}{rgb}{0.00000,0.44700,0.74100}%
\definecolor{mycolor2}{rgb}{0.85000,0.32500,0.09800}%
\definecolor{mycolor3}{rgb}{0.92900,0.69400,0.12500}%
\pgfplotsset{every axis plot/.append style={thick}}
\begin{tikzpicture}
\small
\begin{axis}[%
width=\columnwidth,
height=0.7\columnwidth,
xmin=0,
xmax=90,
ymin=0,
ymax=0.1,
y tick label style={
        /pgf/number format/.cd,
        fixed,
        precision=2,
        /tikz/.cd
    },
axis background/.style={fill=white},
xmajorgrids,
ymajorgrids,
legend style={legend cell align=left, align=left, draw=white!15!black,at={(0.9,0.7)}}
]
\addplot [color=black]
  table[row sep=crcr]{%
1	0.0977901145226596\\
11	0.0973981964551598\\
21	0.0967150980405345\\
31	0.0958174901838474\\
41	0.0947630896000984\\
51	0.0935956813431207\\
61	0.0923486898493108\\
71	0.0910477547569782\\
81	0.0897126128362826\\
91	0.0883584888488305\\
};
\addlegendentry{Exact values using \eqref{Eq:ExpressionForTheBtermBinaomial}}

\addplot [color=black,densely dashdotted]
  table[row sep=crcr]{%
1	0.0109037869638213\\
11	0.0086282150907108\\
21	0.00710292818934062\\
31	0.00601254691850099\\
41	0.00519638214124601\\
51	0.0045639882105104\\
61	0.00406058967483666\\
71	0.00365109561054984\\
81	0.00331200126143324\\
91	0.00302698185008498\\
};
\addlegendentry{CR Bound in \eqref{eq:CR_Binomial_Linear} }

\addplot [color=black,dashed]
  table[row sep=crcr]{%
1	0.00832028867760387\\
11	0.00672576580984685\\
21	0.00564695854457635\\
31	0.00486823566920973\\
41	0.00427950620449019\\
51	0.00381868358215499\\
61	0.00344808946824884\\
71	0.00314353053689295\\
81	0.00288875974994748\\
91	0.00267246054977245\\
};
\addlegendentry{Universal CR bound in \eqref{eq:CR_Universal_Bound_Binomial}}

\end{axis}

\end{tikzpicture}%
%
	\caption{Evaluation of the  exact Bregman risk in \eqref{Eq:ExpressionForTheBtermBinaomial}, the CR bound in \eqref{eq:CR_Binomial_Linear}, and the universal CR bound in \eqref{eq:CR_Universal_Bound_Binomial}. The parameters are set to $\alpha=3,\beta=5$ and $a=1$.  }
	\label{fig:Bregman_CR_Binomial}
		 \vspace{-0.2cm}
\end{figure}
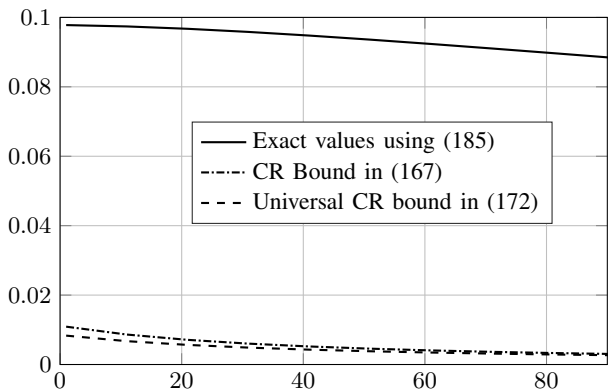

Proposition~\ref{prop:Evaluation_Beta_Bregman} can now be used to asses how tight the CR bound is. 
By combing the CR lower bound in \eqref{eq:CR_Universal_Bound_Binomial} with the upper bound in \eqref{eq:UpperBoundBinoamialBeta}  and using  that $ \lim_{n \to \infty}  c^\star  a n=1$,  and that $d^\star =\Theta(\frac{1}{n})$, we have that  for the beta prior 
\begin{align}
R_{\phi}(X|Y)  =\Theta \left( \frac{1}{n} \right). 
\end{align} 
In other words, the CR bound in  \eqref{eq:CR_Universal_Bound_Binomial}, and, hence, the tighter version in \eqref{eq:CR_Binomial_Linear}  are order tight.  
The CR bound in \eqref{eq:CR_Binomial_Linear}, the universal CR bound in \eqref{eq:CR_Universal_Bound_Binomial}   and the exact expression for the risk in \eqref{Eq:ExpressionForTheBtermBinaomial} are shown in Fig.~\ref{fig:Bregman_CR_Binomial}.

Unfortunately, while the bounds can be show to be order tight, the value of $n$ needs to be rather large before the bounds become effective. One possible way to remedy this is to   tighten the bound in \eqref{eq:SeconBoundOnDelta}, which was used in the proof of the CR bound in \eqref{eq:CR_Binomial_Linear}. In particular, the following tighter version of the bound can be shown 
\begin{align}
 \frac{1}{ \Delta_\phi(u,v)} &\le   2  \E \left[ T_{u,v}(1-T_{u,v})^2 \right] \\
        &=\frac{1}{15}  \hspace{-0.1cm} \left(12 u^3+u^2 (9 v-30) \right. \notag\\
        & \left. +u \left(6 v^2-20 v+20\right)+v \left(3 v^2-10 v+10\right)\right). 
\end{align}
We leave this for the future work. Also note that, to the best of our knowledge, there are no alternative bounds in the literature that could be used as a baseline for comparison.

\section{Conclusion} 
This paper has proposed a general class of Bayesian lower bounds  for the case when the underlying loss function is a  BD.   The approach allows for deriving a version of the CR bound that is specific to a given BD. To show the applicability of the new CR bound it has been evaluated for the Poisson noise case and Binomial noise case. For both examples, the bounds have been shown to admit the same behavior as the corresponding Bregman risk in the  low noise regime, hence, demonstrating the effectiveness of the new CR bound.

\section*{Acknowledgement} 
We would like  to acknowledge Semih Yagli. 
In particular, the scalar version of Lemma~\ref{lem:BregmanRemeinder} was developed together with him. 

\begin{appendices}

\section{Proof of Lemma~\ref{lem:FactsAboutGamma}}
\label{app:lem:FactsAboutGamma}

The expressions for   $\E[X ]$  and $\mathbb{V}(X)$ are standard. 
  To show \eqref{eq:FractionalMOmentGamma} we use the following  well-known integral:
  \begin{align}
      \int_0^\infty x^{k} \eu^{-\alpha x} {\rm d} x= \left \{ \begin{array}{cc} 
 \infty,  & k \le -1, \\
 \frac{\Gamma(k+1)}{\alpha^{k+1}}, & k>-1 .
 \end{array} \right.   \label{eq:ScoreFunctionTypeTerm}
  \end{align}
  Therefore,
  \begin{align}
  \E \left[ \frac{1}{X^n} \right]=  \frac{\alpha^\theta}{ \Gamma(\theta)  }  \int   x^{\theta-1-n} \eu^{-\alpha x} {\rm d} x
  = \alpha^n \frac{ \Gamma(\theta-n)}{ \Gamma(\theta)  }  ,
  \end{align} 
  for $\theta>n$ and infinity otherwise.  

To show \eqref{eq:ScoreFunctionTypeTerm} observe that
\begin{align}
    \rho_X(x)= \frac{\theta-1}{x}-\alpha, \,  x >0 
\end{align}
and, hence,
\begin{align}
\hspace{-0.3cm} \E \left[  \left(\frac{\theta-1}{X}-\alpha \right)^2     \right]& \hspace{-0.05cm} =   \hspace{-0.05cm} \E \left[  \frac{(\theta-1)^2}{X^2}-2\alpha \frac{\theta-1}{X}  +\alpha^2     \right] \\
&  \hspace{-0.05cm}  =  \hspace{-0.05cm}  \left \{  \hspace{-0.2cm} \begin{array}{cc}  
\infty, & \theta \le 2, \\
\frac{(\theta-1) \alpha^2}{(\theta-2)  }- 2 \alpha^2+\alpha^2, & \theta > 2, \end{array}  \right. 
\end{align} 
where we have used that  $\E \left[ \frac{1}{X^2} \right]= \alpha^2 \frac{ \Gamma(\theta-2)}{ \Gamma(\theta)}= \frac{\alpha^2}{(\theta-2) (\theta-1)}$ for $\theta>2$ and $\E \left[ \frac{1}{X} \right]= \alpha \frac{ \Gamma(\theta-1)}{ \Gamma(\theta)}= \frac{\alpha}{\theta-1}$ for $\theta>1$. 
Moreover,  
\begin{align}
\E \left[  \left(\frac{\theta-1}{X}-\alpha \right)^2   X  \right]&=\E \left[  \frac{(\theta-1)^2}{X}- 2 \alpha (\theta-1)+ \alpha^2X      \right]\\
&= \left \{ \begin{array}{cc} 
\infty, & \theta \le 1 \\
\alpha, & \theta > 1
\end{array}  \right.  ,
\end{align}
where have used that $\E \left[ \frac{1}{X} \right]= \alpha \frac{ \Gamma(\theta-1)}{ \Gamma(\theta)}= \frac{\alpha}{\theta-1}$ for $\theta>1$. 

The proof of \eqref{eq:EstimatorForGamma} can be found in \cite{dytso2019estimationPoisson}.
  This concludes the proof.

\section{Proof of Theorem~\ref{thm:CRboundPoisson}}
\label{app:thm:CRboundPoisson}

Using the CR bound in \eqref{eq:CRboundNew}  and the expression for the score function in \eqref{eq:ScoreFunction_Poisson} we have that 
\begin{align}
 \hspace{-0.5cm}  &\! \E \left[ \frac{(  \frac{\rm d}{{\rm d}X} \log f_{YX}(Y,X) )^2}{\Delta_\phi(X,\E[X|Y])} \right]\\
   &= \E \left[ \frac{\left(  \frac{Y}{X}-a+\rho_X(X) \right)^2}{\Delta_\phi(X, c_1 Y+c_2)} \right]\\
    &\le \E \left[ \hspace{-0.05cm} \left(  \frac{Y}{X}-a+\rho_X(X) \right)^2 \hspace{-0.1cm}  \left( \frac{4X}{3} + \frac{2 ( c_1 Y+c_2)}{3} \right) \hspace{-0.05cm}\right] ,\label{eq:BoundOnTiltedFisher_Thm5}
\end{align}
where in the last step we have used the bound in \eqref{eq:BoundOnTheTilt}.  We now compute individual terms  in \eqref{eq:BoundOnTiltedFisher_Thm5}.

The first term in \eqref{eq:BoundOnTiltedFisher_Thm5}  is computed as follows:  
\begin{align}
     &\! \E \left[ \left(  \frac{Y}{X}-a+\rho_X(X) \right)^2  X   \right]  \notag\\
      &=    \E \left[ \left( \frac{Y}{X}-a \right)^2 X   \right]  + 2  \E \left[ \left(\frac{Y}{X}-a \right) \rho(X) X   \right]   \notag\\
      &\quad + \E \left[  \rho_X^2(X) X    \right] \\
       &=    \E \left[ \left( \frac{Y}{X}-a \right)^2 X   \right]   + \E \left[  \rho_X^2(X) X    \right] \\
       &=   a    + \E \left[  \rho_X^2(X) X    \right] , \label{eq:ScoreFunctionandX}
\end{align}
where in \eqref{eq:ScoreFunctionandX} we have used that $\E \left[ \left( Y-aX \right)^2 |  X   \right]=aX$.    

The second term in \eqref{eq:BoundOnTiltedFisher_Thm5} is computed as follows: 
\begin{align}
      &\! \E \left[ \left(  \frac{Y}{X}-a+\rho_X(X) \right)^2  Y  \right]\\
      &= \E \left[ \left(  \frac{Y}{X}-a \right)^2  Y  \right]+ 2\E \left[ \left(  \frac{Y}{X}-a \right) \rho_X(X)  Y  \right] \notag\\
      &\quad +\E \left[   \rho_X^2(X)   Y  \right]\\
      &= \E \left[ \left(  \frac{Y}{X}-a \right)^2  Y  \right]+\E \left[   \rho_X^2(X)   Y  \right]\\
      &= \E \left[ \left(  \frac{Y}{X}-a \right)^2  Y  \right]+a\E \left[   \rho_X^2(X)   X  \right]\\
         &= \E \left[   \frac{Y^3}{X^2}- 2a\frac{Y^2}{X} +a^2 Y  \right]+a\E \left[   \rho_X^2(X)   X  \right]\\
         &= \E \left[   \frac{a^3 X^3+3 a^2 X^2+aX}{X^2}- 2a\frac{aX+a^2X^2}{X} +a^3 X  \right] \notag\\
         &\quad +a\E \left[   \rho_X^2(X)   X  \right]\\
          &= a^2+a \E \left[ \frac{1}{X} \right] +a\E \left[   \rho_X^2(X)   X  \right], \label{eq:ScoreAndY}
\end{align}
where we have used that 
\begin{align}
 \hspace{-0.1cm} \E \left[ \frac{Y}{X} \rho_X(X)  Y  \right]&=  \E \left[ \frac{aX+a^2X^2}{X} \rho_X(X)    \right]=-a^2, \\
    \E \left[  \rho_X(X)  Y  \right]&= \E \left[  \rho_X(X)  aX  \right] =-a.
\end{align}
Furthermore,  the third term has been computed in \eqref{eq:CR-MMSEcomp:eq2} and is given by 
\begin{align}
\E \left[ \left(  \frac{Y}{X}-a+\rho_X(X) \right)^2    \right]   =  \E \left[ \frac{a}{X} \right] +\E \left[ \rho_X^2(X)  \right]. \label{eq:TheLAstTErmINBremgnaPoison}
\end{align}

Finally, combining \eqref{eq:BoundOnTiltedFisher_Thm5}, \eqref{eq:ScoreFunctionandX},  \eqref{eq:ScoreAndY}  and \eqref{eq:TheLAstTErmINBremgnaPoison} leads to  the desired bound. 
 This concludes the proof.

\section{Proof of Proposition~\ref{prop:EvaluationOfbregmanForGamma}}
\label{app:prop:EvaluationOfbregmanForGamma}

The proof of \eqref{eq:LinearAndOptimalRisk} follows from Lemma~\ref{lem:FactsAboutGamma} where we have shown that 
\begin{align}
   \E[X|Y=y]&=   \frac{1}{\alpha+a}y +\frac{\theta}{\alpha+a} = c^\star y+d^\star.
\end{align} 

From the structure of the Bregman divergence we have that 
\begin{align}
R_\phi( X|Y ) &= \E \left[X \log X \right] \notag\\
&\quad  -  \E \left[ X   \log \left ( \frac{Y}{\alpha+a} +\frac{\theta}{\alpha+a} \right) \right]  \notag\\
&\quad-  \E \left[X- \frac{Y}{\alpha+a} -\frac{\theta}{\alpha+a} \right]
\end{align}
Now by using Lemma~\ref{lem:FactsAboutGamma} where we have shown that
\begin{align}
 \E \left[X- \frac{Y}{\alpha+a} -\frac{\theta}{\alpha+a} \right]=  0,
 \end{align}
and
\begin{align}
\E \left[X \log X \right]&=  \frac{\alpha^\theta}{ \Gamma(\theta)  }   \int_0^\infty x \log (x)  x^{\theta-1} \eu^{-\alpha x} {\rm d}x \\
&=    \frac{\theta \left(\log\left(\frac{1}{\alpha}\right)+\psi \left(\theta+1\right)\right)}{\alpha},
\end{align}
where $\psi(t)$ is the  digamma function.  Next, observe that $x \log x$ is a convex function,  and hence, 
\begin{align}
 &\E \left[  X   \log \left ( \frac{Y}{\alpha+a} +\frac{\theta}{\alpha+a} \right) \right] \notag\\
 &= \E \left[ \E[ X|Y]   \log \left ( \frac{Y}{\alpha+a} +\frac{\theta}{\alpha+a} \right) \right] \\
& =\E \left[  \left ( \frac{Y}{\alpha+a} +\frac{\theta}{\alpha+a} \right)   \log \left ( \frac{Y}{\alpha+a} +\frac{\theta}{\alpha+a} \right) \right]  \label{Eq:ExpressionForB}\\
& = \E \left[ \E \left[  \left ( \frac{Y}{\alpha+a} +\frac{\theta}{\alpha+a} \right)   \log \left ( \frac{Y}{\alpha+a} +\frac{\theta}{\alpha+a} \right) |X \right] \right] \\
& \ge   \E \left[   \left ( \frac{ \E \left[Y|X \right] }{\alpha+a} +\frac{\theta}{\alpha+a} \right)   \log \left ( \frac{\E \left[Y|X \right]}{\alpha+a} +\frac{\theta}{\alpha+a} \right) \right] \label{eq:JensenBoundOnBregmean:eq1} \\
& =   \E \left[   \left ( \frac{ aX}{\alpha+a} +\frac{\theta}{\alpha+a} \right)   \log \left ( \frac{aX}{\alpha+a} +\frac{\theta}{\alpha+a} \right) \right] ,
\end{align} 
where in \eqref{eq:JensenBoundOnBregmean:eq1} we have used  Jensen's inequality.  Observe that \eqref{Eq:ExpressionForB} leads to \eqref{Eq:ExpressionForTheBterm}, the fact that $Y$ is according to a negative-binomial was show in \cite{dytso2019estimationPoisson}. 

To show the upper bound in \eqref{eq:BoundsOnB}  observe that
\begin{align}
&\E \left[  X   \log \left ( \frac{Y}{\alpha+a} +\frac{\theta}{\alpha+a} \right) \right] \notag\\
&=  \E \left[  X   \E\left[\log \left ( \frac{Y}{\alpha+a} +\frac{\theta}{\alpha+a} \right)|X \right] \right]\\
& \le  \E \left[  X  \log \left (  \frac{ \E\left[ Y |X \right] }{\alpha+a} +\frac{\theta}{\alpha+a} \right)\right]\\
& =  \E \left[  X  \log \left (  \frac{ aX  }{\alpha+a} +\frac{\theta}{\alpha+a} \right)\right],
\end{align}
where the last step is a consequence of Jensen's inequality.  
This concludes the proof.

\section{Proof of Lemma~\ref{lem:Beta_Quantities}}
\label{app:lem:Beta_Quantities}

The proof of \eqref{eq:Term_HPfunction}  follows by using the Euler integral representation of a hypergeometric function \cite[Ch.~15.3]{abramowitz1965handbook}. 

The score function is found by differentiating \eqref{eq:pdf_Beta} and is  given by 
\begin{align}
\rho_X(x)&=\frac{\alpha-1}{x}-\frac{1-\beta}{1-x}, \, x\in (0,1). 
\end{align}
Moreover,  for $k\in \mathbb{R}$
\begin{align}
\E[\rho_X^2(X) X^k ]&=  (\alpha-1)^2 \E \left[ \frac{X^k }{X^2} \right] +(1-\beta)^2 \E \left[ \frac{X^k }{(1-X)^2} \right] \notag\\
&+2 (1-\alpha)(1-\beta) \E \left [ \frac{X^k }{X (1-X) } \right]\\
&=(\alpha-1)^2   \frac{c_{\alpha,\beta}}{c_{\alpha-2+k,\beta}}+ (1-\beta)^2   \frac{c_{\alpha,\beta}}{c_{\alpha+k,\beta-2}} \notag\\
&+2 (1-\alpha)(1-\beta)   \frac{c_{\alpha,\beta}}{c_{\alpha+k-1,\beta-1}} . 
\end{align}

Since the beta distribution is a conjugate prior for a binomial noise \cite{diaconis1979conjugate}, the conditional expectation is a linear function of $y$. Now, if the conditional expectation is a linear function, than it can only be of  of the  form given in  \cite{ericson1969note} by
\begin{align}
\E[ X|Y=y]=\frac{ y  \mathbb{V}(X)  +   \E[  X ] \E[  X (1-aX)  ]}{  na   \mathbb{V}(X)  +  \E[ X (1-aX) ] }. 
\end{align}

\section{Proof of Theorem~\ref{thm:CR_for_Linear_Estimators_Binomial}}
\label{app:thm:CR_for_Linear_Estimators_Binomial}

       Let   the polynomial in the \eqref{eq:SeconBoundOnDelta} be denoted by 
            \begin{align}
            \psi(u,v)&=  \frac{4u}{3} + \frac{2v}{3} - u^2- \frac{ 2uv}{3}-\frac{v^2}{3}. 
            \end{align}
            Moreover,
            \begin{align}
             \psi(X,c_1Y+c_2)&= - \frac{c_1^2 Y^2}{3} + \frac{2c_1(1-c_2) Y}{3}   - \frac{2c_1 X Y  }{3} \notag\\
             &\quad -X^2+\frac{2(2-c_2)X}{3}+\frac{2c_2-c_2^2}{3}.
            \end{align}

            Using the expression of the score function in  \eqref{eq:ScoreFunctionBinomial+Input} and the bound in \eqref{eq:SeconBoundOnDelta}, we have that 
            \begin{align}
   &\! \E \left[ \frac{(  \frac{\rm d}{{\rm d}X} \log f_{YX}(Y,X) )^2}{\Delta_\phi(X, c_1Y+c_2)} \right]\\
   &= \E \left[ \frac{\left(  \frac{an X-Y}{aX^2-X}+\rho_X(X) \right)^2}{\Delta_\phi(X, c_1Y+c_2)} \right]\\
       & \le   \E \left[ \left(   \frac{an X-Y}{aX^2-X}+\rho_X(X) \right)^2   \psi(X,c_1Y+c_2)\right] \\
       & = \sum_{0 \le r \le 2, 0 \le m \le 2}   \kappa_{r,m}  G_X(a;r,m), 
\end{align}
where  we have defined
\begin{align}
G_X(a;r,m)&= \E \left[ \left(   \frac{an X-Y}{aX^2-X}+\rho_X(X) \right)^2  X^r Y^m    \right]  \\
 &=  \E \left[ \left(   \frac{an X-Y}{aX^2-X} \right)^2 X^r Y^m \right]  \notag\\
 &+  \E[\rho_X^2(X) X^r Y^m  ]   \notag\\
& +2 \E \left[ \left(   \frac{an X-Y}{aX^2-X} \right)\rho_X(X) X^r Y^m \right]. \label{eq:ExpressionForGFunction}
\end{align} 

We will also need the following expression for the moments of the binomial distribution \cite{knoblauch2008closed}: 
\begin{align}
\E[ Y^m |X  ]= \sum_{k=0}^m   (n)_k  s_{m,k}   (aX)^k, \label{eq:binomial_moments}
\end{align} 
where  $s_{k,m}$ is  Striling's number of the second kind and $(n)_k$ is the falling factorial. 

Next, we individually compute each term of \eqref{eq:ExpressionForGFunction}.
To compute the first term of \eqref{eq:ExpressionForGFunction} we will need the following expression: 
 \begin{align}
& \E[ (anX-Y)^2 Y^m| X]  \\\
  &=  \E[ ((anX)^2 -2an XY + Y^2 )Y^m| X] \\
 &=(anX)^2 \E[ Y^m |X  ]-2 an X \E[ Y^{m+1} |X  ]+ \E[ Y^{m+2} |X  ]\\
 &=  (anX)^2  \sum_{k=0}^m    (n)_k  s_{m,k}  (aX)^k \notag\\ 
 &- 2 an X  \sum_{k=0}^{m+1}  (n)_k  s_{m+1,k}    (aX)^k 
  + \sum_{k=0}^{m+2}    (n)_k s_{m+2,k}   (aX)^k \label{eq:applying_binomial_moments}\\
 &= \sum_{k=0}^{m+2}  d_{k,m,n}  (aX)^k, \label{eq:Conditiona_Variance_Almost}
 \end{align} 
 where \eqref{eq:applying_binomial_moments} follows by using \eqref{eq:binomial_moments}; and in \eqref{eq:Conditiona_Variance_Almost} we have defined 
 \begin{align}
&d_{k,m,n} \notag\\
&=  n^2   (n)_{k-2} s_{m,k-2}  - 2 n   (n)_{k-1} s_{m+1,k-1}+ (n)_k s_{m+2,k}  , 
 \end{align} 
with $s_{m,-1} =s_{m,-2}=0$.  
 
 Therefore, the first term in \eqref{eq:ExpressionForGFunction} is given by 
 \begin{align}
&  \E \left[ \left(   \frac{an X-Y}{aX^2-X} \right)^2 X^r Y^m \right] \notag\\
  &=    \E \left[ \frac{ \E[\left( an X-Y \right)^2 Y^m  | X]}{  \left(  aX^2-X \right)^2}  X^r \right]\\
    &=    \E \left[ \frac { \sum_{k=0}^{m+2}  d_{k,m,n} (aX)^k }{  \left(  aX^2-X \right)^2}  X^r \right]\\
       &=  \sum_{k=0}^{m+2}  d_{k,m,n} a^k    \E \left[ \frac {   X^{k+r-2} }{  \left( 1 -a X \right)^2}  \right].  \label{eq:EvaluationOf_the_first_Term}
 \end{align}

 The second term in \eqref{eq:ExpressionForGFunction} is given by
 \begin{align}
  \E[\rho_X^2(X) X^r Y^m  ]&=    \E[\rho_X^2(X) X^r \E[Y^m|X]  ]\\
  &=   \E[\rho_X^2(X) X^r  \sum_{k=0}^m  (n)_k s_{m,k}     (aX)^k  ]  \label{eq:applying_binomial_moments_2}\\
  &=  \sum_{k=0}^m  (n)_k  s_{m,k}   a^k  \E[\rho_X^2(X) X^{k+r} ], \label{eq:EvaluationOf_the_second_Term}
 \end{align} 
 where \eqref{eq:applying_binomial_moments_2} follows by using \eqref{eq:applying_binomial_moments}. 
 
 To compute the third term in \eqref{eq:ExpressionForGFunction}  we will need the following expression: 
 \begin{align}
& \E[ (anX-Y) Y^m| X] \notag\\
 &=  an X \E[Y^m|X]- \E[Y^{m+1}|X]\\
 &= anX  \sum_{k=0}^m   (n)_k s_{m,k}     (aX)^k  - \sum_{k=0}^{m+1}  (n)_k s_{m+1,k}     (aX)^k \label{eq:applying_binomial_moments_3}\\
 &= \sum_{k=0}^{m+1}  h_{k,m,n}   (aX)^k, \label{eq:Definition_h}
 \end{align} 
where in \eqref{eq:applying_binomial_moments_3} we have used \eqref{eq:applying_binomial_moments}; and in \eqref{eq:Definition_h} we  
have defined
 \begin{align}
 h_{k,m,n}  =  n (n)_{k-1}   s_{m,k-1}    -    (n)_k s_{m+1,k}. 
 \end{align}
Now, the third term in \eqref{eq:ExpressionForGFunction} can be simplified to
 \begin{align}
 & \E \left[ \left(   \frac{an X-Y}{aX^2-X} \right)\rho_X(X) X^r Y^m \right] \notag\\
  &=   \E \left[ \left(   \frac{\E[ (anX-Y) Y^m| X] }{aX^2-X} \right)\rho_X(X) X^r \right]\\
  &=   \E \left[ \left(   \frac{  \sum_{k=0}^{m+1}  h_{k,m,n}   (aX)^k }{aX^2-X} \right)\rho_X(X) X^r \right] \label{eq:Definition_h_apply}\\
    &=   \sum_{k=0}^{m+1}  h_{k,m,n}  a^k   \E \left[  \frac{   X^{k+r-1} }{aX-1}  \rho_X(X) \right] \\
     &=   \sum_{k=0}^{m+1}  h_{k,m,n}  a^k   \E \left[  - \frac{{\rm d}}{{\rm d} X} \frac{   X^{k+r-1} }{aX-1}  \right]  \label{eq:diffrentiability_scoreFunction}\\
          &=  \sum_{k=0}^{m+1}  h_{k,m,n} a^{k+1}   \E \left[   \frac{ X^{k+r-1} }{( 1- aX )^2}  \right] \notag\\
           &+   \sum_{k=0}^{m+1}  h_{k,m,n} a^{k} (k+r-1) \E \left[   \frac{ X^{k+r-2} }{1-aX}  \right] , \label{eq:applying_Derivatie_binomial}
\end{align}
where  \eqref{eq:Definition_h_apply}  follows by using  \eqref{eq:Definition_h}; \eqref{eq:diffrentiability_scoreFunction} follows by using the property of the score function that $\E[ g(X) \rho_X(X)]= - \E[   \frac{{\rm d}}{{\rm d} X} g(X)]$ for any function $g$;  and \eqref{eq:applying_Derivatie_binomial} follows by using that
\begin{align}
- \frac{{\rm d} \left(\frac{x^{k+r-1}}{a x-1}\right)}{ {\rm d}  x}=\frac{a x^{k+r-1}}{(1-a x)^2}+\frac{(k+r-1) x^{k+r-2}}{1-a x}. 
\end{align}

Combining \eqref{eq:ExpressionForGFunction} with evaluations in  \eqref{eq:EvaluationOf_the_first_Term}, \eqref{eq:EvaluationOf_the_second_Term}, \eqref{eq:applying_Derivatie_binomial} concludes the proof.

\section{Proof of Proposition~\ref{prop:Evaluation_Beta_Bregman}}
\label{app:prop:Evaluation_Beta_Bregman}
To show \eqref{eq:LinearAndOptimalRiskBinomial} observe that, in view of Lemma~\ref{lem:Beta_Quantities}, it holds for the beta distribution that
\begin{align}
\E[X|Y=y]=c^\star y+d^\star. 
\end{align}
That is, the optimal estimator is linear.  

The condition for validity of the CR bound has already been verified in the proof of Proposition~\ref{prop:EvaluationForBetaMMSE}. 

To show the exact expression for $R_{\phi}(X|Y) $,  first, note that 
\begin{align}
R_{\phi}(X|Y) &=   \E \left[ X \log \frac{X(1-\E[X|Y])}{\E[X|Y](1-X)} - \frac{X - \E[X|Y]}{1-\E[X|Y]}  \right]\\
&= \  \E \left[ X \log \frac{X(1-\E[X|Y])}{\E[X|Y](1-X)}  \right] \label{eq:errorTermZero}\\
&= \E\left[X \log  \frac{X}{1-X}\right]   + B,
\end{align}
where  in \eqref{eq:errorTermZero} we have used that $\E \left[ \frac{X - \E[X|Y]}{1-\E[X|Y]}  \right]=0$.  Second, the term $B$ can be expressed as 
\begin{align}
B&=\E \left[ X \log \frac{1-\E[X|Y]}{\E[X|Y]}  \right]\\
&= \E \left[ \E[X|Y] \log \frac{1-\E[X|Y]}{\E[X|Y]}  \right]\\
&= \E \left[ ( c^\star Y+d^\star) \log \frac{1-c^\star Y-d^\star }{ c^\star Y+d^\star }  \right],
\end{align}
where in the last step we have used \eqref{eq:CE_Beta_Distribution}. 

To find the distribution of $Y$ observe that
\begin{align}
P_Y(y)&=  {{n} \choose {y} }  \E\left[    (aX)^y (1-aX)^{n-y}    \right] \notag\\
&= {{n} \choose {y} }  \frac{c_{\alpha,\beta}  a^y     F_{2,1}(\alpha+y,y-n;\beta+\alpha+y;a ) }{  c_{\alpha+y,\beta} } ,
\end{align}
where the last expression follows from \eqref{eq:Term_HPfunction}.

The logarithmic moments of the beta distribution have been computed in \cite{ebrahimi1999ordering}  and are given by 
\begin{align}
 \E\left[X \log  X\right] &=   \frac{\alpha ( \psi(\alpha+1) -\psi(\alpha+\beta+1)  )}{\alpha+\beta} ,\\
  \E\left[X \log  (1-X)\right] &=   \frac{\alpha ( \psi(\beta) -\psi(\alpha+\beta+1) ) }{\alpha+\beta} .
 \end{align} 
 Therefore,
 \begin{align}
  \E\left[X \log  \frac{X}{1-X}\right] = \frac{\alpha}{\alpha+\beta} ( \psi(\alpha+1)-\psi(\beta) )  .
 \end{align} 
 
 Finally, to upper bound the term $B$, we use  the fact that the function $u \to  u \log \frac{1-u}{u}$ is  concave  together with  Jensen's inequality to arrive at
 \begin{align}
 B &= \E \left[ \E\left[ ( c^\star Y+d^\star) \log \frac{1-c^\star Y-d^\star }{ c^\star Y+d^\star }   | Y \right] \right]\\
 & \le \E \left[ ( c^\star \E[Y|X]+d^\star) \log \frac{1-c^\star  \E[Y|X]-d^\star }{ c^\star  \E[Y|X]+d^\star }   \right]\\
  & = \E \left[ ( c^\star anX +d^\star) \log \frac{1-c^\star  anX-d^\star }{ c^\star  anX+d^\star }   \right].
 \end{align}
 This concludes the proof. 

\end{appendices}

\bibliography{refs}

\begin{thebibliography}{10}
\providecommand{\url}[1]{#1}
\csname url@samestyle\endcsname
\providecommand{\newblock}{\relax}
\providecommand{\bibinfo}[2]{#2}
\providecommand{\BIBentrySTDinterwordspacing}{\spaceskip=0pt\relax}
\providecommand{\BIBentryALTinterwordstretchfactor}{4}
\providecommand{\BIBentryALTinterwordspacing}{\spaceskip=\fontdimen2\font plus
\BIBentryALTinterwordstretchfactor\fontdimen3\font minus
  \fontdimen4\font\relax}
\providecommand{\BIBforeignlanguage}[2]{{%
\expandafter\ifx\csname l@#1\endcsname\relax
\typeout{** WARNING: IEEEtran.bst: No hyphenation pattern has been}%
\typeout{** loaded for the language `#1'. Using the pattern for}%
\typeout{** the default language instead.}%
\else
\language=\csname l@#1\endcsname
\fi
#2}}
\providecommand{\BIBdecl}{\relax}
\BIBdecl

\bibitem{weinstein1988general}
E.~Weinstein and A.~J. Weiss, ``A general class of lower bounds in parameter
  estimation,'' \emph{IEEE Trans. Inf. Theory}, vol.~34, no.~2, pp. 338--342,
  1988.

\bibitem{van2004detection}
H.~L. Van~Trees, \emph{Detection, Estimation, and Modulation Theory, Part {I}:
  {D}etection, Estimation, and Linear Modulation Theory}.\hskip 1em plus 0.5em
  minus 0.4em\relax John Wiley \& Sons, 2004.

\bibitem{ziv1969some}
J.~Ziv and M.~Zakai, ``Some lower bounds on signal parameter estimation,''
  \emph{IEEE Trans. Inf. Theory}, vol.~15, no.~3, pp. 386--391, 1969.

\bibitem{MMSEKLpreparation}
A.~{Dytso}, M.~{Fau\ss}, A.~M. {Zoubir}, and H.~V. {Poor}, ``{MMSE} bounds for
  additive noise channels under {K}ullback-{L}eibler divergence constraints on
  the input distribution,'' \emph{IEEE Trans. Signal Process.}, vol.~67,
  no.~24, pp. 6352--6367, 2019.

\bibitem{MMSEJointArxiv}
M.~{Fau\ss}, A.~{Dytso}, and H.~V. {Poor}, ``{MMSE} bounds under
  {K}ullback-{L}eibler divergence constraints on the joint input-output
  distribution,'' \emph{arXiv:2006.03722}, 2020.

\bibitem{bregman1967relaxation}
L.~M. Bregman, ``The relaxation method of finding the common point of convex
  sets and its application to the solution of problems in convex programming,''
  \emph{USSR Comput. Math. \& Math. Phys.}, vol.~7, no.~3, pp. 200--217, 1967.

\bibitem{frigyik2008functional}
B.~A. Frigyik, S.~Srivastava, and M.~R. Gupta, ``Functional {B}regman
  divergence and {B}ayesian estimation of distributions,'' \emph{IEEE Trans.
  Inf. Theory}, vol.~54, no.~11, pp. 5130--5139, 2008.

\bibitem{iyer2012submodular}
R.~Iyer and J.~A. Bilmes, ``Submodular-{B}regman and the {L}ov{\'a}sz-{B}regman
  divergences with applications,'' in \emph{Adv. Neural Inf. Process Syst.},
  2012, pp. 2933--2941.

\bibitem{wang2014bregman}
L.~Wang, D.~E. Carlson, M.~R. Rodrigues, R.~Calderbank, and L.~Carin, ``A
  {B}regman matrix and the gradient of mutual information for vector {P}oisson
  and {G}aussian channels,'' \emph{IEEE Trans. Inf. Theory}, vol.~60, no.~5,
  pp. 2611--2629, 2014.

\bibitem{csiszar1991least}
I.~Csiszar, ``Why least squares and maximum entropy? {A}n axiomatic approach to
  inference for linear inverse problems,'' \emph{Ann. Statist.}, vol.~19,
  no.~4, pp. 2032--2066, 1991.

\bibitem{banerjee2005clustering}
A.~Banerjee, S.~Merugu, I.~S. Dhillon, and J.~Ghosh, ``Clustering with
  {B}regman divergences,'' \emph{J. Mach. Learn. Res.}, vol.~6, pp. 1705--1749,
  2005.

\bibitem{banerjee2005optimality}
A.~Banerjee, X.~Guo, and H.~Wang, ``On the optimality of conditional
  expectation as a {B}regman predictor,'' \emph{IEEE Trans. Inf. Theory},
  vol.~51, no.~7, pp. 2664--2669, 2005.

\bibitem{I-MMSE}
D.~Guo, S.~Shamai, and S.~Verd{\'u}, ``Mutual information and minimum
  mean-square error in {G}aussian channels,'' \emph{IEEE Trans. Inf. Theory},
  vol.~51, no.~4, pp. 1261--1282, 2005.

\bibitem{palomar2005gradient}
D.~P. Palomar and S.~Verd{\'u}, ``Gradient of mutual information in linear
  vector gaussian channels,'' \emph{IEEE Trans. Inf. Theory}, vol.~52, no.~1,
  pp. 141--154, 2005.

\bibitem{guo2008mutual}
D.~Guo, S.~Shamai, and S.~Verd{\'u}, ``Mutual information and conditional mean
  estimation in {P}oisson channels,'' \emph{IEEE Trans. Inf. Theory}, vol.~54,
  no.~5, pp. 1837--1849, 2008.

\bibitem{atar2012mutual}
R.~Atar and T.~Weissman, ``Mutual information, relative entropy, and estimation
  in the {P}oisson channel,'' \emph{IEEE Trans. Inf. Theory}, vol.~58, no.~3,
  pp. 1302--1318, 2012.

\bibitem{taborda2014information}
C.~G. Taborda, D.~Guo, and F.~Perez-Cruz, ``Information-estimation
  relationships over binomial and negative binomial models,'' \emph{IEEE Trans.
  Inf. Theory}, vol.~60, no.~5, pp. 2630--2646, 2014.

\bibitem{jiao2018mutual}
J.~Jiao, K.~Venkat, and T.~Weissman, ``Mutual information, relative entropy and
  estimation error in semi-martingale channels,'' \emph{IEEE Trans. Inf.
  Theory}, vol.~64, no.~10, pp. 6662--6671, 2018.

\bibitem{folland2005higher}
\BIBentryALTinterwordspacing
G.~B. Folland, ``Higher-order derivatives and {T}aylor's formula in several
  variables.'' [Online]. Available:
  \url{\url{https://sites.math.washington.edu/\textasciitilde{}folland/Math425/taylor2.pdf}}
\BIBentrySTDinterwordspacing

\bibitem{li2012fast}
L.~Li, G.~Lebanon, and H.~Park, ``Fast {B}regman divergence {NMF} using
  {T}aylor expansion and coordinate descent,'' in \emph{Proceedings of the 18th
  ACM SIGKDD International Conference on Knowledge Discovery and Data Mining},
  2012, pp. 307--315.

\bibitem{gordon1962quantum}
J.~P. Gordon, ``Quantum effects in communications systems,'' \emph{Proc. IRE},
  vol.~50, no.~9, pp. 1898--1908, 1962.

\bibitem{shamai1990capacity}
S.~S. Shamai, ``Capacity of a pulse amplitude modulated direct detection photon
  channel,'' \emph{Proc. Inst. Electr. Eng.}, vol. 137, no.~6, pp. 424--430,
  1990.

\bibitem{diaconis1979conjugate}
P.~Diaconis and D.~Ylvisaker, ``Conjugate priors for exponential families,''
  \emph{Ann. Statist.}, pp. 269--281, 1979.

\bibitem{dytso2019estimationPoisson}
A.~Dytso and H.~V. Poor, ``Estimation in {P}oisson noise: {P}roperties of the
  conditional mean estimator,'' \emph{arXiv preprint arXiv:1911.03744}, 2019.

\bibitem{abramowitz1965handbook}
M.~Abramowitz and I.~A. Stegun, \emph{Handbook of Mathematical Functions with
  Formulas, Graphs, and Mathematical Table}.\hskip 1em plus 0.5em minus
  0.4em\relax National Bureau of Standards Applied Mathematics series 55, 1965.

\bibitem{ericson1969note}
W.~A. Ericson, ``A note on the posterior mean of a population mean,'' \emph{J.
  R. Stat. Soc. B}, vol.~31, no.~2, pp. 332--334, 1969.

\bibitem{knoblauch2008closed}
A.~Knoblauch, ``Closed-form expressions for the moments of the binomial
  probability distribution,'' \emph{{SIAM} J. Appl. Math.}, vol.~69, no.~1, pp.
  197--204, 2008.

\bibitem{ebrahimi1999ordering}
N.~Ebrahimi, E.~Maasoumi, and E.~S. Soofi, ``Ordering univariate distributions
  by entropy and variance,'' \emph{J. Econometrics}, vol.~90, no.~2, pp.
  317--336, 1999.

\end{thebibliography}
\bibliographystyle{IEEEtran}
\end{document}